\documentclass{article}

\usepackage{PRIMEarxiv}

\usepackage[utf8]{inputenc} 
\usepackage[T1]{fontenc}    
\usepackage{hyperref}       
\usepackage{url}            
\usepackage{booktabs}       
\usepackage{amsfonts}       
\usepackage{nicefrac}       
\usepackage{microtype}      
\usepackage{lipsum}
\usepackage{fancyhdr}       
\usepackage{graphicx}       
\graphicspath{{media/}}     

\usepackage[utf8]{inputenc}
\usepackage{amsmath}
\usepackage{amsfonts}
\usepackage{amssymb}
\usepackage{mathrsfs}  
\usepackage{mathtools}
\usepackage{graphicx}
\usepackage{xcolor}
\usepackage{algorithmic}
\usepackage{algorithm}
\usepackage{geometry}
\usepackage{wasysym}
\usepackage{url}
\usepackage{float}
\usepackage{relsize}
\usepackage[makeroom]{cancel}
\usepackage{marginnote}
\usepackage{eucal}
\usepackage{comment}
\newtheorem{remark}{Remark}

\newtheorem{proposition}{Proposition}

\makeatletter      
\renewcommand{\ps@plain}{%
	\renewcommand{\@oddhead}{\hfil\textrm{\thepage}}%
	\renewcommand{\@evenhead}{\@oddhead}%
	\renewcommand{\@oddfoot}{}
	\renewcommand{\@evenfoot}{\@oddfoot}}
\makeatother     

\newcommand{\R}{ \ensuremath{\mathbb{R}}}  

\DeclarePairedDelimiterX\set[1]{\lbrace}{\rbrace}{  #1 }  

\DeclarePairedDelimiterX\norm[1]{\lVert}{\rVert}{#1}  			
\DeclarePairedDelimiterX\inner[2]{(}{)}{ #1\, , #2}  	

\providecommand{\bo}{\mathbf}

\providecommand{\diag}{\mathrm{diag}}
\providecommand{\cov}{\mathrm{COV}}

\DeclareMathSymbol{\Mu}{\mathalpha}{operators}{"4D}

\title{Numerical considerations and a new implementation for ICS
}

\author{Aurore Archimbaud\thanks{Erasmus School of Economics, 
Erasmus University Rotterdam (\texttt{archimbaud@ese.eur.nl}, ORCID: 0000-0002-6511-9091).}
\and 
Zlatko Drma\v{c}\thanks{Department of Mathematics, Faculty of Science,  University of Zagreb, (\texttt{zlatko.drmac@math.hr}, ORCID: 0000-0001-6845-332X).}
\and
Klaus Nordhausen\thanks{Department of Mathematics and Statistics, University of Jyv\"askyl\"a (\texttt{klaus.k.nordhausen@jyu.fi}, ORCID: 0000-0002-3758-8501).}
\and 
Una Radoji\v{c}i\'{c}\thanks{Institute of Statistics \& Mathematical Methods in Economics, Vienna University of Technology (\texttt{una.radojicic@tuwien.ac.at}, ORCID: 0000-0003-0329-0595).} 
\and 
Anne Ruiz-Gazen\thanks{Toulouse School of Economics, Universit\'e  Toulouse 1 Capitole (\texttt{anne.ruiz-gazen@tse-fr.eu}, ORCID: 0000-0001-8970-8061).}}

\begin{document}
\maketitle

\begin{abstract}
Invariant Coordinate Selection (ICS) is a multivariate data transformation and a dimension reduction method that can be useful in many different contexts. It can be used for outlier detection or cluster identification, and can be seen as an independent component or a non-Gaussian component analysis method. The usual implementation of ICS is based on a joint diagonalization of two scatter matrices, and may be numerically unstable in some ill-conditioned situations. We focus on one-step M-scatter matrices and propose a new implementation of ICS based on a pivoted QR factorization of the centered data set. This factorization avoids the direct computation of the scatter matrices and their inverse and brings numerical stability to the algorithm. Furthermore, the row and column pivoting leads to a rank revealing procedure that allows computation of ICS when the scatter matrices are not full rank. Several artificial and real data sets illustrate the interest of using the new implementation compared to the original one.
\end{abstract}


\section{Introduction}\label{S=Intro}
At the heart of multivariate statistics is the investigation of the relationship between the different components of a $p$-variate dataset $X_n=(x_1,\ldots,x_n) \in \R^{p\times n}$.
Traditionally this is first investigated using the empirical covariance matrix
\begin{equation}\label{eq:cov}
\cov(X_n)= \frac{1}{n-1} \sum_{i=1}^n (x_i-\bar{x}_n)( x_i-\bar{x}_n)^\top
\end{equation}
where $\bar{x}_n= 1/n\sum_{i=1}^n x_i$ is the empirical mean and $^\top$ denotes the transpose operator.

Many alternatives for the covariance matrix, known as scatter matrices, were suggested in the literature. Basically, any positive semi-definite matrix-valued function $S$ is a scatter matrix if it is  affine equivariant in the sense that
\[
S(A X_n +  b e^\top) = A S(X_n) A^\top,
\]
where $A$ is a full rank $p \times p$ matrix, $b$ a $p$-vector and $e$ a vector of ones with dimension determined by the context.

It can then be shown that if $x_i$, $i=1,\ldots,n$, is a sample from an elliptical distribution, all scatter matrices are proportional to each other at the population level, if they exist. And as traditional multivariate statistics assumes that the sample follows a Gaussian or an elliptical distribution, analysis of one scatter matrix is sufficient when analyzing 
the sample, as all scatter matrices carry essentially the same information.  
In recent years, interest increased in investigating data that might not follow an elliptical model, in which case it might be of interest to compare two scatter matrices. One such comparison is nowadays established as Invariant Coordinate Selection (ICS) (see \cite{Tyler2009} and \cite{NordhausenRuizGazen2021}), that uses a generalized eigenvalue decomposition for the comparison. However, in the ICS literature, the computation of ICS is usually not discussed. The topic of this article is how to compute ICS in an efficient and numerically stable way for some combinations of scatter matrices that have proven their value in several contexts such as outlier detection \cite{archimbaud2018CSDA}, independent component analysis (ICA) \cite{NordhausenTyler2015} or principal axis analysis (PAA) \cite{CritchleyPiresAmado2006}. 

Many scatter matrices were suggested in the literature where the computation of most of them involves using iterative and approximate algorithms. The one-step M-scatter estimators are weighted covariance matrices where the weights are a function of the Mahalanobis distance. They can be defined explicitly and exactly. However, the Mahalanobis distance involves
the inverse symmetric positive definite matrix $\cov^{-1}$. This inverse is usually computed using the spectral decomposition of $\cov$.
When using $\cov^{-1}$, one tacitly assumes that $\cov$ is numerically positive definite ($\cov\succ 0$ in the L\"{o}wner partial order). 
The same holds true when instead of $\cov$ one uses the triangular Cholesky factor.
Of course, the data could be such that such an assumption is not warranted, and, even if $\cov$ is regular it can be close to the boundary of the cone of the positive definite matrices and thus numerically ill-conditioned. 
A numerical algorithm should detect such ill-conditioning and determine a numerical rank of the matrix. 

To illustrate the point, take $n=3$ centered observations in $p=2$ dimensions:
\begin{equation}\label{eq:Example:2x3}
X_n= \begin{pmatrix}
1-\alpha & 1+\alpha & -2 \cr 1 & 1 & -2
\end{pmatrix},
\end{equation}
where $\alpha$ is small, $|\alpha| \ll 1$; say $\alpha=10^{-8}$. 
The covariance matrix is
\begin{equation}\label{eq:cov:2x3}
\cov(X_n) =  \frac{1}{2}\begin{pmatrix}
6+2\alpha^2 & 6 \cr 6 & 6
\end{pmatrix}=\begin{pmatrix} 3+\alpha^2 & 3 \cr 3 & 3\end{pmatrix}.
\end{equation}
Then $\cov$ can be made exactly singular with a perturbation of size $10^{-16}$ -- it suffices to add $-\alpha^2$ to the position $(1,1)$. 
In fact, in a standard double precision (64 bit) IEEE floating point arithmetic, $\cov$ is computed and stored as exactly singular matrix with all entries equal $3$ (stored as binary number in the corresponding working precision). Hence, in this case, computation involving the inverse $\cov^{-1}$ must fail. 
Furthermore, computing $\cov^{-1}$ from a spectral decomposition of $\cov$ is numerically ill-conditioned: small eigenvalues can be computed entirely wrong and their inverses appear as dominating in $\cov^{-1}$. In fact, in extreme cases a numerical method may compute the smallest eigenvalues even with a wrong sign yielding a failure of the computation of $\cov^{-1}$. Similarly, if we attempt to compute the Cholesky factor of the matrix $\cov$ in (\ref{eq:Example:2x3}), the R or the Matlab function \texttt{chol(.)} returns an error.

On the other hand, it can be checked that the smallest singular value of $X_n$ in (\ref{eq:Example:2x3}) is $O(|\alpha|)$. In this case, it takes a perturbation of at least $10^{-8}$ in the spectral norm to make $X_n$ rank deficient. 
The numerically computed singular values of $X_n$ are accurate up to nearly eight decimal digits -- squaring them we obtain the eigenvalues of $\cov$ also up to eight correct digits -- this was not possible by computing them directly from $\cov$.
The numerical rank deficiency indicates near linear dependence, and that critical decision is better to be made based on the data matrix $X_n$ and the Eckart-Young-Mirsky theorem (see \cite{eckart1936approximation} and \cite{mirsky1960symmetric}), before using $\cov$. In the present paper, we propose a new implementation of ICS in the case of one-step M-scatter matrices. This new algorithm is based on a QR transformation of the centered data. This transformation allows to bypass completely the computation of the inverse or the inverse square root of the covariance matrix that are needed in the usual implementation of ICS with one-step M-estimators. This algorithm is able to cope with data sets that are approximately multicollinear. The numerical stability of the algorithm is also ensured by 
row and column pivoting. Moreover, we introduce a rank revealing procedure that is useful when data are multicollinear.

The structure of this paper is as follows. In  Section~\ref{sec::ICS} we will discuss ICS in detail, including its usual implementation. In Section \ref{S=D1}, we will detail the new implementation of ICS using the QR transformation of the centered data. This section includes the derivation  of the different computation steps, the column pivoting
that can be useful to define a dimension reduction procedure, the row pivoting than improves the numerical stability of the algorithm, and a discussion on the computational cost. Section \ref{S=appli} illustrates the new algorithm with applications to clustering, Independent Component Analysis and outlier detection. Finally, Section \ref{S=conclu} concludes the paper.

\section{Invariant coordinate selection} \label{sec::ICS}
ICS can be useful in quite many different contexts. It can be used for example for model selection \cite{KankainenTaskinenOja2007,Tyler2009,NordhausenOjaOllila2011b,Loperfido2021} and can find  Fisher's linear discriminant in an elliptical mixture model without knowing the cluster labels. 
Furthermore,  for instance \cite{archimbaud2018CSDA,archimbaud2018Rjournal} use it for outlier detection as large eigenvalues can be seen as an indicator for anomalies (see \cite{Tyler2009} for details). In the context of independent component analysis (ICA) and non-Gaussian component analysis (NGCA) ICS can be seen as an ICA and NGCA method respectively \cite{Oja2006,NordhausenOjaOllila2008,NordhausenOjaTylerVirta2017,Radojicic2020ngcaTest,NordhausenOjaTyler2022}, if the scatter matrices involved have some additional properties. The main idea in the applications of ICS is to perform ICS and proceed with the analysis of the data using the obtained interesting components, called invariant coordinates.

\textcolor{black}{As discussed,} ICS explores data by comparing two scatter matrices. 
Let us first recall in Section \ref{sec::scatter} the definition of a scatter matrix and the class of scatter matrices we will use in the present paper. Then, 
we briefly recall in Section \ref{SS=pca} two methods based on a single scatter functional: principal component analysis (PCA) and whitening. Finally, 
we discuss ICS and some computational aspects of ICS in Section \ref{SS=ics}.

\subsection{Scatter matrices} \label{sec::scatter}

Let us assume that we have a sample of $n$ $p$-variate observations, $ x_1,\ldots,  x_n$ collected into the $p \times n$ data matrix $ X_n = (x_1,\ldots, x_n)$. 
As mentioned above, a scatter functional is any matrix-variate function that is positive semi-definite and affine equivariant.

The statistical literature is full of different suggestions, where most were made by the robust statistics community, on how to make inference in elliptical models more efficient for heavy-tailed distributions and more resistant against outliers. Based on their robustness properties, which are measured by their influence functions (IF) and breakdown points (BP), \cite{Tyler2009} divide scatter functionals into three categories.
In Class I are scatter functionals with unbounded IF and a BP of essentially zero. Class II has a bounded IF with a BP $\in (0,1/(p+1)]$. Class III has a bounded IF and a BP $> 1/(p+1)$. In the following, we will focus on Class I estimators only.
Class I includes the usual covariance functional but also the following one-step M-estimators with a functional defined by:

\[
\cov_{w}(X_n)=\frac{1}{n} \sum_{i=1}^n w(D^2(x_i))(x_i - \bar{ x}_n)(x_i - \bar{ x}_n)^\top,
\]
where $D^2(x_i)=(x_i - \bar{x}_n)^\top \cov(X_n) ^{-1} (x_i - \bar{x}_n)$ is the squared Mahalanobis distance and $w$ is a non-negative and continuous weight function. For convenience, in the following, we will often drop the dependence of $\cov$ and $\cov_w$ on $X_n$.

The covariance matrix is obtained using $w(d)=1$ (up to the factor $1/(n-1)$ instead of $1/n$), and we get the $\cov_{-1}$ matrix defined by \cite{CritchleyPiresAmado2006} when $w(d)=1/d$. As noticed by \cite{NordhausenTyler2015}, when $w(d)=d^{\alpha}$ with $\alpha<0$, such estimators downweight values with large Mahalanobis distance and have a robust flavor even if they have a zero breakdown point.
The fourth-moment based estimator $\cov_4$ is also a Class I estimator, obtained with $w(d)=d$. It is highly nonrobust since it upweights values with large Mahalanobis distances but it proves to be useful in particular situations as detailed below.\\

\subsection{Principal component analysis and whitening}\label{SS=pca}

PCA is a method that searches a new representation of the data in which the coordinates are uncorrelated and the ordering of the components is according to their relevance. The measure of relevance used in PCA is variation and therefore the first component should have maximal variation, the $k$th component should have the $k$th largest variation under the constraint of being uncorrelated to the previous $k-1$ components ($k=2,\ldots,p$). Classical PCA uses the regular variance to measure variation and principal components can be obtained 
using the eigenvector-eigenvalue decomposition of the covariance matrix, i.e.,
$$
\cov(X_n)= U( X_n) D( X_n) U(X_n)^\top,
$$
where the orthogonal $p \times p$ matrix $U(X_n)$ contains the eigenvectors of $\cov( X_n)$ as its columns, and $D(X_n)$ is a $p \times p$ diagonal matrix containing the corresponding eigenvalues in an ascending order on its diagonal. The principal components are then
$$
z_i(X_n) = U(X_n)^\top ( x_i - \bar{ x}_n),\quad i=1,\ldots,n.
$$
PCA is in detail discussed for example in \cite{Jolliffe2002}. 

Now, as all scatter functionals are proportional for elliptical data, provided they exist (see for example \cite{NordhausenTyler2015}), then all scatter matrices will have the same eigenvectors in the same order, and the difference when doing the eigenvector-eigenvalue decomposition is only in the magnitude of the eigenvalues which measure the variation in the sense of the 
scatter functional used. We can also replace $\bar{x}_n$ in PCA with another affine equivariant location estimator $T(X_n)$ and do the rotation using the eigenvectors of any scatter matrix $S(X_n)$.

While PCA has many useful properties, a big drawback is that it is not affine equivariant, meaning that for $X_n^* = A  X_n$
$$
z_i^*(X_n^*)  = U(X_n^*)^\top ( x_i^* -  T(X_n^*)) = U( X_n)^\top (x_i - T(X_n)) = z_i
$$
will in general not be true for $A$ being a full rank $p \times p$ matrix. It will  only be true if $A$ is an orthogonal matrix, provided we are not interested in the signs of the components.\\


Whitening is a data transformation that goes one step further than PCA. Aside from giving centered and uncorrelated components, 
whitening also removes any scale differences. The transformation is defined as 
\begin{equation}\label{zd:xist}
x_i^{st} = S(X_n)^{-1/2}(x_i - T(X_n)), i=1,\ldots,n, 
\end{equation}
and therefore has the properties
$$
T(X_n^{st}) = 0 \quad \mbox{and} \quad S(X_n^{st}) = I_p,
$$
where $X_n^{st}=(x_1^{st},\ldots,x_n^{st})$ and $S(X_n)^{1/2}$ is the unique positive definite square root of $S(X_n)$.

However, as \cite{IlmonenOjaSerfling2012} pointed out, there are at least four ways on how to replace the inverse square root $S(X_n)^{-1/2}$ in (\ref{zd:xist}), and these all differ by a rotation. 
Therefore, the representation $X_n^{st}$ is not unique. However, all variants remove the information from the data that is measured using the estimators $T$ and $S$, and in the case of elliptical data, the standardized components are then spherically distributed.

\subsection{ICS and computational aspects of ICS}\label{SS=ics}

The general idea of invariant coordinate selection (ICS) is to combine whitening with PCA using a different scatter matrix for each transformation. 
It is well established that there is no best scatter matrix combination for ICS and the best choice depends on the data and the purpose. \cite{Tyler2009}  argue mainly against using two scatter functionals from Class III. 
Although no scatter combination is considered best, there is one combination sticking out, the so-called fourth order blind identification (FOBI) combination \cite{Cardoso1989} uses $\cov$ and $\cov_4$,
and is for example reviewed in \cite{NordhausenVirta2019}.
FOBI is highly non-robust but can for example be used in an ICA and NGCA context. Its main advantage is that 
it is moment-based and therefore many of its properties are easily derived. In the present paper we consider scatter pairs of the form $\cov$-$\cov_w$ which include FOBI but also the principal axis analysis \cite{CritchleyPiresAmado2006}.
If we take the two scatter matrices $\cov$ and $\cov_w$, then the ICS (unmixing) matrix $B(X_n)$ is the matrix that jointly diagonalizes $\cov$ and $\cov_w$, i.e.,
\begin{equation}\label{eq:eqICS}
B(X_n) \cov\, B(X_n)^\top = I_p \quad \mbox{and} \quad B(X_n) \cov_w \, B( X_n)^\top = D( X_n),
\end{equation}
where $D(X_n)$ is a diagonal matrix with diagonal elements in decreasing order. For a location estimator $T(X_n)$, the invariant coordinates are then defined as
$$
z_i = B(X_n) \left(x_i - T(X_n) \right), \ i=1,\ldots,n.
$$

This transformation was first denoted generalized PCA \cite{Caussinus1990InterestingPO,Caussi2007PCAclustering,CAUSSINUS2003237}, but is nowadays better known as ICS due to the following invariance property
\[
B(X_n)A^{-1} = J B(A X_n +   b e^\top)
\]
provided the diagonal elements in $ D$ are all distinct, where $A$ is a $p\times p$ matrix and $b$ a $p$-vector. $J$ denotes a sign change matrix, i.e., a diagonal matrix with $\pm 1$ on its diagonal. Thus, for the invariant coordinates, we have
\[
z_i(X_n)=B(X_n)\left(x_i - T(X_n) \right) = J B(A X_n  + b e^\top)\left((A x_i +b) - T( AX_n + b e^\top) \right),
\]
which means that the invariant coordinates  $z_i$ are affine invariant under linear transformations up to their signs. It is then often argued that ICS finds the intrinsic structure of the data. 

If however not all diagonal elements in $D$ are distinct, then still the 
ordering of the diagonal elements in $D(A X_n + b e^\top )$ is as in $D(X_n)$, and the space corresponding to the unique 
eigenvalues remains invariant. More precisely, assume $d_1\geq\dots\geq d_m$ are the distinct diagonal elements in $D(X_n)$ with multiplicities $p_1,\dots,p_m$, $\sum p_i=p$, and let $z(X_n)=(z_{(1)}(X_n),\dots, z_{(m)}(X_n))$ be the corresponding partition of the invariant components. Then, if $p_i=1$, $z_{(i)}( X_n)$ and $z_{(i)}(A X_n+ b e^\top)$ are equal up to a sign. On the other hand, if $p_i>1$, then the components in $z_{(i)}(X_n)$ and $z_{(i)}(A X_n+ b e^\top)$ span the same space. It is worth mentioning that in methods like non-Gaussian component analysis (NGCA) and non-Gaussian independent component analysis (NGICA), the goal is to identify the non-Gaussian subspace of the data, and thus separate the meaningful signal from the Gaussian noise. The identification of the signal subspace is then often done using two-scatter estimators satisfying certain properties~\cite{Tyler2009,Oja2006}, where the common approach is to discard as noise the components belonging to the equal eigenvalues. For more details see e.g.~\cite{Radojicic2020ngcaTest}.   

For convenience, if the context is clear, we will in the following often drop the dependence on $X_n$ not only for $\cov$ and  $\cov_w$, but also for other matrices as $B$ and $D$.

The basic interpretation of ICS is that $\cov$ is used for whitening the data and then a PCA using $\cov_w$ is applied to the whitened data to see if $\cov_w$ can still find any structure not yet removed by $\cov$.
The eigenvalues in $D(X_n)$ and the eigenvectors in $B(X_n)$ can also be derived through the spectral decomposition of the following symmetric matrix:
\begin{equation}\label{eq:symics}
M(X_n)= \cov^{-1/2} \cov_w \cov^{-1/2} = U D U^\top.    
\end{equation}
with $U$ a $p\times p$ orthogonal matrix such that $B=U^\top\cov^{-1/2}$.

For the purpose of this paper fixing the signs of the invariant coordinates is not relevant. However \cite{ICS,NordhausenOjaOllila2011b} discuss ways to address this issue, for example by using a second location estimator $T_2(Z_n)$ and fixing the signs of the columns of $B$ such that all components of $T_2(Z_n)$ are positive. A general discussion on the role of location estimators in the context of ICS can be found in \cite{AlashwaliKent2016}.


An issue we have left open so far is the computation of the matrices $D$ and $ B$. And this is in fact not much discussed in the literature, where usually it is just stated that it can be formulated as the generalized eigenvalue-eigenvector problem. However, in several recent applications 
it was realized 
that the computation for complex data requires some further thoughts. 


Algorithm \ref{ALG:2:ICS2} details the usual implementation of ICS for the scatter pair $\cov$-$\cov_w$ when the data are preliminary  centered (using usually $\bar{x}_n$). It is based on two spectral decompositions. First, the eigenvalues-eigenvectors of $\cov$ are computed in order to derive the inverse $\cov^{-1}$ (needed to compute the Mahalanobis distances and $\cov_w$) and the inverse square root $\cov^{-1/2}$ (needed to compute $M(X_n)$). Then the eigenvalues-eigenvectors of $M(X_n)$ are computed and the invariant coordinates (or components) are derived.
As explained in Section \ref{S=Intro} and illustrated in Section \ref{S=appli}, such an algorithm is not numerically stable as soon as the data are ill-conditioned. 
In what follows, we propose a new implementation of ICS that avoids the spectral decomposition of $\cov$ and will solve these numerical instability issues.

\begin{algorithm}[ht]
	\caption{$(D_2, \, B,\, Z) = \textsf{ICSEigen}(X_n^c)$}
	\label{ALG:2:ICS2}
	\begin{algorithmic}[1]
		\REQUIRE Data $X_n^c \in\R^{p\times n}$, $n>p$
		\STATE Compute $\cov$ and $\cov_w$;
		\STATE Compute the eigenvalue - eigenvector decomposition of $ \cov$: $\cov = {U}_1  D_1 U_1^\top$;
		\STATE Compute the symmetric inverse square root of $\cov$: 
		${ \cov^{-1/2}} =  U_1 D_1^{-1/2} U_1^\top$;
		\STATE Compute $M(X_n)$:
			$M(X_n) = \cov^{-1/2} \cov_w \cov^{-1/2}$;
		\STATE Compute the eigenvalue - eigenvector decomposition of $ M(X_n)$: $M(X_n) = U_2 D_2 U_2^\top$;
		\STATE Compute $ B$: $B = U_2^\top \cov^{-1/2}$ \COMMENT{The signs of $B$ can be fixed.};
		\STATE Compute $Z$: $Z =  B  X_n^{c}$.  \COMMENT{The signs of $Z$ can be fixed.};
		\ENSURE  $\mbox{diag}(D_2) \in \R^p$, $B\in\R^{p\times p}$, $Z\in\R^{p\times n}$ 
	\end{algorithmic}
\end{algorithm}

\section{A new implementation of ICS}\label{S=D1}

The classical implementation of ICS described above for the scatter pairs $\cov$-$\cov_w$ relies on the spectral decomposition of $\cov$ and the computation of its inverse and its inverse square root. These computations are prone to numerical instability (see Section~\ref{S=appli}).
In Section~\ref{S=Intro}, we preliminary discussed an advantage of using positive definite and semidefinite matrices implicitly, through their natural factorizations that are already available with explicitly given factors in their very definitions, such as e.g. the definition of $\cov$ through $X_n$ in \eqref{eq:cov}. This \emph{natural factor formulation} is an important technique for solving ill-conditioned problems in e.g. finite element computation \cite{argyris-naturalFEM-75}, \cite{vavasis-96-FEMWC} and solving Lyapunov equations \cite{ham-82}. 

In Section~\ref{SS=Implicit_Cholesky}, we provide  matrix computation details  that are the key ingredients of a more stable procedure. 
In Section~\ref{SS=Implicit_Cholesky} we show their usefulness in the computational framework of ICS.
In Section~\ref{SS=QRCP_review} we review the numerical details of the rank revealing pivoted QR factorization. In Section~\ref{SS=cost}, we give the algorithm we propose and discuss its computational cost.


\subsection{Using the implicitly computed Cholesky factor}\label{SS=Implicit_Cholesky}
Here we assume that $n> p$, or even $n\gg p$, so that the centered data matrix $X_n^c=X_n - \bar{x}_n e^\top$ is short and wide matrix,
$$
X_n^c = \begin{pmatrix} 
* & * & * & *   \cr
* & * & * & *   \cr
\end{pmatrix}\in\R^{p\times n},\;\; n\gg p.
$$
Let 
\begin{equation}\label{eq:QR_hatX_T}
\Pi_2^\top \left(\frac{1}{\sqrt{n -1}}X_n^{c\top}\right)\Pi_1 =  Q  R = \begin{pmatrix} 
* & *    \cr
* & *    \cr
* & *    \cr
* & *    \cr
\end{pmatrix}
\begin{pmatrix} 
\bullet & \bullet   \cr
0 & \star    \cr
\end{pmatrix},\;\; Q \in\R^{n\times p},\;\; Q ^\top Q =I_p,
\end{equation}
be the QR factorization of $X_n^c$ with optional row and column pivoting.\footnote{This is the ``short", or economy size QR factorization of a (typically) tall and skinny matrix. Further, the scaling factor $1/\sqrt{n-1}$ in (\ref{eq:QR_hatX_T}) is immaterial for the factorization as it can be applied afterwards, whenever necessary. Such details are taken care of in a software implementation.}
Pivoting means that we can reorder the observation $x_i$ using the permutation encoded in the permutation matrix $\Pi_2$, and the coordinates $1,\ldots, p$ according to the permutation matrix $\Pi_1$. 
\begin{remark}
Since $X_n^ce=\mathbf{0}$, we see that $e^\top  Q  R =\mathbf{0}$, so that in the case of nonsingular $ R $ it must hold that $e^\top Q =\mathbf{0}$.
\end{remark}

To ease the notation, let us take the permutations to be identities -- in fact if we redefine $X_n^c$ by initially reordering its rows and columns, then we can remove $\Pi_1$, $\Pi_2$ from the notation ($X_n^c\equiv \Pi_2^\top X_n^{c\top}\Pi_1$) and when we reinterpret the result take into account that all has been permuted.   

Then
$$
\cov = \frac{1}{n-1}X_n^cX_n^{c\top} =  R ^\top  Q ^\top Q  R  =  R ^\top R  = \cov^{1/2} \cov^{1/2} .
$$
If $R$ is nonsingular and we set $W = \cov^{1/2} R^{-1}\in\R^{p\times p}$, then it follows that $W^\top W = I_p$, i.e. $\cov^{1/2} = W  R $. The matrix $ R $ is a triangular factor of $\cov$ -- if we fix the signs of the diagonals as positive, this factor is the uniquely determined Cholesky factor. Clearly, any matrix of the form $W R $, with arbitrary orthogonal $W$ can be used as a ``square root" of $\cov$ (see also \cite{IlmonenOjaSerfling2012}). The non-uniqueness of a particular ``square root" 
carried by the orthogonal matrix $W$ is immaterial in  context ICS   \cite{IlmonenOjaSerfling2012} as the subsequent rotation is affine equivariant under orthogonal transformations.

Also, if $ R  = U\Sigma V^\top$ is the SVD of $ R $ (an economy size SVD of $X_n^{c\top}$ is $X_n^{c\top} = ( Q U) \Sigma V^\top$), then $\cov=V\Sigma^2 V^\top$,
$\cov^{1/2} = V \Sigma V^\top$, and $\cov^{-1/2} = V \Sigma^{-1} V^\top$.
Hence, when $\cov^{1/2}$ or its inverse are needed, both can be computed from the SVD of $X_n^c$ or $ R $. This is certainly numerically more stable and it may give reasonably accurate results when the eigendecomposition of the explicitly computed $\cov$ fails; recall the discussion in Section~\ref{S=Intro}.

In the next proposition, we show how the QR factorisation of $X_n^{c\top}$ allows for an efficient computation of the Mahalanobis distance and $\cov_w$, without using $\cov$.

\begin{proposition}\label{PROP:S2x}
\mbox{ }
\begin{itemize}
    \item[(i)] The Mahalanobis distances 
$$
D^2(x_i)=(x_i-\bar{x}_n)^\top \cov^{-1}(x_i-\bar{x}_n)
$$ 
can be computed as $D^2(x_i)=(n-1)q_i$,
where $q_i = \| Q (i,:)\|_2^2$, $i=1,\ldots, n$,  are the statistical leverage scores of the rows of $X_n^{c\top}$ \cite[Definition 1]{Drineas-Mahony-2012}.
\item[(ii)] The matrix $\cov_w$ can be written as
\begin{equation}\label{eq:S2x-QQT}
\cov_w = \frac{1}{n} X_n^c\;\mathrm{Diag}(w((n-1)q_i))_{i=1}^n X_n^{c\top}.
\end{equation}
\end{itemize}
\end{proposition}
{\sc Proof}: 
\begin{itemize}
    \item[(i)] The key observation is that
$$
X_n^{c\top} \cov^{-1}X_n^c = \sqrt{n-1} Q  R  R^{-1} (R ^{-1})^\top  \sqrt{n-1}R ^\top  Q ^\top = 
(n-1)  Q  Q ^\top,
$$
where $ Q  Q ^\top$ is an orthogonal projector. 
 \item[(ii)]
From the definition of $\cov_w$ we have 
\begin{eqnarray*}
\cov_w&=&\frac{1}{n}\sum_{i=1}^n w((x_i-\bar{x}_n))^\top \cov^{-1}(x_i-\bar{x}_n))) (x_i -\bar{x}_n))(x_i -\bar{x}_n))^\top \\ 	
&=& \frac{1}{n} X_n^c\; \mathrm{Diag}(w(\mathrm{diag}(X_n^{c\top} \cov^{-1}X_n^c))) \; X_n^{c\top},
\end{eqnarray*}
\hfill $\blacksquare$
\end{itemize}

\subsection{Computing $M(X_n)$}\label{SS=S2y}
Consider now $M(X_n) = \cov^{-1/2} \cov_w \cov^{-1/2}$. 
Going back to $\cov^{1/2} = W  R $, we have (because of symmetry) 
$\cov^{1/2} = W  R  = R ^\top W^\top$ and  $\cov^{-1/2} =  R^{-1}W^\top = 
W  R ^{-T}$.
Hence (keeping in mind the representation of $\cov_w$ in Proposition \ref{PROP:S2x})
$$
X_n^{c\top} \cov^{-1/2} = \sqrt{n-1} Q  R  R^{-1}W^\top = \sqrt{n-1} Q W^\top,
$$
$$
\cov^{-1/2}X_n^c = (X_n^{c\top} \cov^{-1/2})^\top = \sqrt{n-1} W  Q ^\top,
$$
and
\begin{equation}\label{eq:S2y-W}
M(X_n) = \cov^{-1/2} \cov_w \cov^{-1/2} = \frac{n-1}{n} W  Q ^\top \; \mathrm{Diag}(w((n-1)q_i))_{i=1}^n \;  Q W^\top .
\end{equation}
 The non-uniqueness of the choice of the square root of $\cov$  carries over into $M(X_n)$. 
If we continue to use the Cholesky factor $ R $, we set $W=I_p$ (the identity matrix) and consider the particular choice
$$
{\widetilde{M}(X_n)} = \frac{n-1}{n}   Q ^\top \; \mathrm{Diag}(w((n-1)q_i))_{i=1}^n \;  Q .
$$
Let 
\begin{equation}\label{eq:eig(S2yhat)}
\widetilde{M}(X_n) = \widetilde{U}_2 \widetilde{D}_2 \widetilde{U}_2^\top,\;\mbox{ and } \;M(X_n) = U_2 D_2 U_2^\top
\end{equation}
be the spectral decompositions ($\widetilde{D}_2$, $D_2$ diagonal; $\widetilde{U}_2$, $U_2$ orthogonal). 
Since $M(X_n) = W \widetilde{M}(X_n) W^\top$ is orthogonal similarity, if we assume that the eigenvalues in both $D_2$ and $\widetilde{D}_2$ are decreasingly ordered, then $\widetilde{D}_2=D_2$.
Hence, any function of the eigenvalues of $M(X_n)$ can be computed based solely on the matrix $ Q $ from the QR factorization (\ref{eq:QR_hatX_T}).
Note that the spectral decomposition (\ref{eq:eig(S2yhat)}) can be obtained implicitly from the SVD of $\mathrm{Diag}(\sqrt{w((n-1)q_i)})_{i=1}^n  Q $ -- this guarantees (implicitly) both the symmetry and the (semi)definiteness.

The situation with the eigenvectors is different, because they depend on the particular choice of $W$. Since 
$$
M(X_n) = W \widetilde{M}(X_n) W^\top = W \widetilde{U}_2 \widetilde{D}_2 \widetilde{U}_2^\top W^\top,
$$
we can take $U_2 = W \widetilde{U}_2$. That we do not know $W$ is immaterial\footnote{
Here $W$ represents a global change of coordinates in $\R^p$ by a ``rotation".} because we use $U_2$ to compute
$$
B = U_2^\top \cov^{-1/2} = \widetilde{U}_2^\top W^\top W  R ^{-T} = \widetilde{U}_2^\top  R ^{-T}.
$$

As usual with ICS, $\widetilde{U}_2$ is not uniquely determined. Let the eigenvalues of $M(X_n)$ appear, say, in the decreasing order on the diagonal of $\widetilde{D}_2$ so that multiple eigenvalues 
occupy successive positions. If there are $\ell$ different eigenvalues with multiplicities 
$m_1,\ldots,m_{\ell}$, then $\widetilde{U}_2$ is determined up to a post-multiplication with a block diagonal
matrix $\varPsi=\Psi_1\oplus \ldots \oplus \Psi_\ell$, where each $\Psi_i$ is arbitrary $m_i\times m_i$  orthogonal matrix. Assume that all eigenvalues are simple. Then $\Psi_i=\pm 1$ for each $i$. When we call a black-box function to compute eigenvalues and eigenvectors numerically, we have no control over those scaling factors. Hence, even in the simplest case of simple eigenvalues,  $B$ is determined up to a pre-multiplication by $\mathrm{diag}(\pm 1)$.   In the case of multiple eigenvalues, we can replace $\widetilde{U}_2$ with $\widetilde{U_2}\varPsi$ and $B$ changes into
$$
\left(\begin{smallmatrix}\Psi_1^\top & & \cr 
& \ddots &   \cr   &   & \Psi_{\ell}^\top\end{smallmatrix}\right) B .
$$


\begin{remark}
A technical remark regarding software implementation is in order. Eigenvalue solvers for real symmetric matrices return the eigenvalues usually ordered into a nondecreasing or nonincreasing sequence, so the computed $D_2$ and $\widetilde{D}_2$ (and the corresponding eigenvector matrices $U_2$, $\widetilde{U}_2$) will be consistently ordered. On the other hand, the computed matrices $M(X_n)$ and $\widetilde{M}(X_n)$ might be computed as non-symmetric and the eigensolver for non-symmetric matrices will be invoked -- as a result, the ordering of the eigenvalues is not a priori known. Hence, if one wants to check the above formulas numerically, the discrepancies may be in the signs but also in the order. This, of course, can be resolved by symmetrizing the computed matrices in an obvious way.
\end{remark}



\subsection{Review of the pivoted QR factorization}\label{SS=QRCP_review}
Now we go back to the pivoted QR factorization (\ref{eq:QR_hatX_T}) and discuss the choice of pivoting matrices $\Pi_1$ and $\Pi_2$. What we have in mind is numerical stability in cases when the full rank assumption that $\cov$ is regular fails or numerically fails (there is a redundancy or near redundancy in the supplied information and $X_n$ is highly ill-conditioned to the extent that it must be treated as rank deficient.)
We recall that pivoting is allowed transformation because it merely reorders the data. If we insist on the initial ordering, it is enough to redefine $ R = R \Pi_1^\top$ and update all formulas above; same with $\Pi_2 Q $. Also, note that the scaling factor $1/\sqrt{n-1}$ has no essential role for the pivoting; in a software implementation we can omit it in the factorization and later adjust the formulas.  

\subsubsection{Rank revealing column pivoting}\label{SS=RRpiv}
Let us first discuss the column pivoting encoded in $\Pi_1$. Suppose we use the Businger-Golub column pivoting \cite{bus-gol-65} which computes
\begin{equation}\label{QRCP}
\left(\frac{1}{\sqrt{n-1}}X_n^{c\top}\right) \Pi_1 =  Q    R ,\;\;\mbox{where}\;\;| R _{ii}|\geq \sqrt{\sum_{k=i}^j
	| R _{kj}|^2},\;\;\mbox{for all}\;\;1\leq i\leq j \leq p.
\end{equation}
Note that in particular $| R _{11}|\geq | R _{22}| \geq \cdots\geq | R _{pp}|$. 
The permutation matrix $\Pi_1$ is determined dynamically. Let $Y^{(1)}=\frac{1}{\sqrt{n-1}}X_n^{c\top}$. In a $k$--th elimination step we first identify  the smallest index $j_k$ such that 
\begin{equation}\label{eq:QRCP-jk}
j_k = \mathrm{argmax}_{j\geq k} \|Y^{(k)}(k:n,j)\|_2 ,
\end{equation}
and then apply the permutation matrix $P_k$ that swaps the $k$th and the $j_k$th column of $Y^{(k)}$ thus giving $Y^{(k)}P_k=({y}_1^{(k)},\ldots,{y}_p^{(k)})$. Then $(Y^{(k)}P_k)(k:n,k:p)$ is transformed by a  $(n-k+1)\times (n-k+1)$ Householder reflector -- the standard step in the QR factorization.

An advantage of this pivoting is that $| R _{ii}|$, $i=1,\ldots, p$, usually mimic the distribution of the singular values of $R$ (i.e. of $X_n^{c\top}$). So, if the matrix is numerically close to being rank $q<p$, we will see that\footnote{This mimics the relation between the $(q+1)$st and the $q$th singular value of $X_n^c$.} $| R _{q+1,q+1}|\ll | R _{qq}|$. In practice, we will set a tolerance $\epsilon>0$ (say, $\epsilon=10^{-8}$, but in general choosing the threshold should depend on the noise level in the data and other factors) and scan along the diagonal until reaching the first index $q$ such that 
$| R _{q+1,q+1}| < \epsilon | R _{qq}|$ (or $| R _{q+1,q+1}| < \epsilon | R _{11}|$, depending on the allowed perturbation or what is the noise level in the data). 

For a statement on the ill--conditioning, we must have an a priori information on the type of uncertainty in the data (see Section~\ref{S=appli} for more details).


Now, if we introduce a partition 
$$
 R  = \begin{pmatrix}  R _{[11]} &  R _{[12]}\cr 
\mathbf{0} &  R _{[22]}\end{pmatrix},\;\;  R _{[11]}\in\R^{q\times q};\;\;
 Q =\begin{pmatrix} Q _{[1]} &  Q _{[2]} \end{pmatrix},\;\;
	 Q _{[1]}\in\R^{n\times q},
$$
then 
\begin{eqnarray*}
\frac{1}{\sqrt{n-1}}X_n^{c\top}\Pi_1 &=& \begin{pmatrix} Q _{[1]} &  Q _{[2]} \end{pmatrix} \begin{pmatrix}  R _{[11]} &  R _{[12]}\cr 
\mathbf{0} &  R _{[22]}\end{pmatrix}\\
&=& \begin{pmatrix}  Q _{[1]} R _{[11]}, &   Q _{[1]} R _{[12]} +  Q _{[2]} R _{[22]}\end{pmatrix}
\end{eqnarray*}
and (because of (\ref{QRCP})) 
$$
\| R _{[22]}\|_F \leq \sqrt{p-q} | R _{q+1,q+1}| < \sqrt{p-q} \epsilon | R _{qq}| \leq \sqrt{p-q} \epsilon | R _{11}|.
$$
If we decide to neglect $R_{[22]}$ and replace it with zero, then we can justify it by saying that we have introduced a perturbation (backward error) into certain $p-q$ columns of $X_n^{c\top}$, and the total size of that perturbation is $\| Q _{[2]} R _{[22]}\|_F=\| R _{[22]}\|_F$. 
More precisely, this reads
$$
\left(\frac{1}{\sqrt{n-1}}X_n^{c\top}
- \begin{pmatrix} \mathbf{0} &    Q _{[2]} R _{[22]}\end{pmatrix}\Pi_1^\top\right)\Pi_1
= 
\begin{pmatrix}  Q _{[1]} R _{[11]}, &   Q _{[1]} R _{[12]} \end{pmatrix}.
$$
This relation precisely identifies which columns of $X_n^{c\top}$ are identified as numerically linearly independent and which columns can be changed with controlled amount of error into linearly dependent on the former ones, to yield rank $q$. 
Then we have, transposing the data matrix back to its original shape, 
$$
\begin{pmatrix} 
* & * & * & *   \cr
* & * & * & *   \cr
\end{pmatrix} = 
\frac{1}{\sqrt{n-1}}\Pi_1^\top X_n^c \approx \begin{pmatrix} 
 R _{[11]}^\top \cr  R _{[12]}^\top\end{pmatrix}  Q _{[1]}^\top=
\begin{pmatrix}\bullet\cr\bullet\end{pmatrix}\begin{pmatrix}
* & * & * & *
\end{pmatrix}.
$$
\begin{remark}
Set
$$
\widetilde{X}_n^\top = 
\left(X_n^{c\top}
- \sqrt{n-1}\begin{pmatrix} \mathbf{0} &    Q _{[2]} R _{[22]}\end{pmatrix}\Pi_1^\top\right).
$$
Then $e^\top \widetilde{X}_n^\top=\mathbf{0}$, and 
$$
\frac{1}{\sqrt{n-1}}\widetilde{X}_n^\top\Pi_1 =  Q _{[1]} \begin{pmatrix}  R _{[11]}, &   R _{[12]} \end{pmatrix},\;\;\;
\frac{1}{\sqrt{n-1}}(\widetilde{X}_n^\top\Pi_1)(:,1:q) =  Q _{[1]}  R _{[11]} .
$$
\end{remark}

\subsubsection{Dimension reduction}\label{SS=dim_red}
Hence, with small (controlled) perturbation, the data is placed in a $q$ dimensional subspace, by keeping the $q$ coordinates that the pivoting $\Pi_1$ selected upfront.  


$$
\begin{pmatrix} 
* & * & * & *   \cr
\end{pmatrix} = 
\frac{1}{\sqrt{n-1}}(\Pi_1^\top X_n^c)(1:q,:) \approx \
 R _{[11]}^\top  Q _{[1]}^\top=
\begin{pmatrix}\bullet\end{pmatrix}\begin{pmatrix}
* & * & * & *
\end{pmatrix}.
$$
This selects the $q$ original observables, places all data in $\R^q$, and the corresponding $q\times q$ covariance matrix is positive definite submatrix of the the matrix $\cov$.

Another way, instead of truncating, is to apply an additional orthogonal  transformation
based on the
QR factorization (can also be done with pivoting $\Pi_3$)
$$
\begin{pmatrix} 
 R _{[11]}^\top \cr  R _{[12]}^\top\end{pmatrix}\Pi_3 = \Omega \begin{pmatrix} T\cr \mathbf{0}\end{pmatrix} = \Omega_1 T,\;\;\Omega = \begin{pmatrix}\Omega_1 & \Omega_2\end{pmatrix}.
$$ 
These two steps correspond to the so called URV decomposition (see \cite{stewart1993determining}).
The new representation of the $q$-dimensional data, in the basis given by $\Omega_1$ is 
$\Pi_1^\top X_n^c \approx \sqrt{n-1} \Omega_1 T\Pi_3^\top  Q _{[1]}^\top$. We keep the basis $\Omega_1$ fixed and the data that  we enter to the next step is $$\widetilde{X}_n = \sqrt{n-1}\, T\,\Pi_3^\top  Q _{[1]}^\top.$$ $\widetilde{X}_n$ is $q\times n$ and there will be no problem with the numerical
definiteness of $\widetilde{X}_n\widetilde{X}_n^\top$.

\subsubsection{On the row pivoting}\label{SS=row_piv}
The final remark in this part of the discussion is on $\Pi_2$ in (\ref{eq:QR_hatX_T}). It relates to the Powell-Reid  pivoting (see \cite{pow-rei-68}). Without going into details, it can be replaced by something simpler and almost equally good. Before the QR factorization with the Businger-Golub column pivoting, the matrix rows (in our case, the rows of $X_n^{c\top}/\sqrt{n-1}$, but here $\sqrt{n-1}$ is irrelevant) should be reordered (by $\Pi_2$) so that their $\ell_\infty$ norms are decreasing. This may help reducing the errors of the finite precision arithmetic. For details see \cite{cox-hig-98}.



\subsubsection{Algorithm and computational cost}\label{SS=cost}

In the full rank case, the new implementation of ICS is summarized in Algorithm \ref{ALG:1:QR-full-rank}. As a consequence of column pivoting in Step 1, the matrix $ R $ is diagonally dominant which makes the computation of $ R^{-1}\widetilde{U}_2$ in Step 5 numerically more accurate. The row pivoting in Step 1 is optional, and it is used only if the rows of $X_n^c$ vary in norm over several orders of magnitude. The QR Decomposition with row an column pivoting can be obtained using the \emph{qr} function in R with LAPACK = TRUE that makes uses of the routines: DGEQP3 and ZGEQP3.

\begin{algorithm}[ht]
	\caption{($D_2$, $B$, $Z$) = \textsf{ICSQR}$(X_n^c)$}
	\label{ALG:1:QR-full-rank}
	\begin{algorithmic}[1]
		%
		
		\REQUIRE $X_n^c\in\R^{p\times n}$, $n>p$. 
		\STATE Compute the pivoted QR factorization $\Pi_2^\top (\frac{1}{\sqrt{n -1}}X_n^{c\top})\Pi_1 =  Q  R $
		\COMMENT{\emph{The row pivoting with $\Pi_2$ is optional. See Section~\ref{SS=row_piv}.}}
		\STATE{To accommodate pivoting, redefine (implicitly) $ Q =\Pi_2 Q $, $ R = R \Pi_1^\top$.}
		\STATE Compute the statistical leverage scores $q_i= \| Q (i,:)\|_2^2$, $i=1,\ldots,n$.
		\STATE Compute the $p$ singular values $\mbox{diag}(D_2)$ and the $p\times p$ right singular vector matrix $\widetilde{U}_2$ of 
		$\mbox{Diag}\left(\sqrt{w((n-1)q_i)}\right)_{i=1}^n Q $.
		\COMMENT{\emph{This represents implicit computation of the eigenvalues $\mbox{diag}(D_2)$ and the
		eigenvector matrix $\widetilde{U}_2$ of 
		$$
		\widetilde{M}(X_n)=\frac{n-1}{n}   Q ^\top \; \mathrm{Diag}(w((n-1)q_i))_{i=1}^n \;  Q 
		$$
		}}
		\STATE Compute $B= ( R^{-1} \widetilde{U}_2)^\top$
		\STATE Compute $Z^\top=\sqrt{n-1} Q \widetilde{U}_2$. \COMMENT{Note that$Z^\top=X_n^{c\top}\Pi_1 B^\top$.}
		\ENSURE $\mbox{diag}(D_2)\in \R^p$, $B\in\R^{p\times p}$, $Z\in\R^{p\times n}$
	\end{algorithmic}
\end{algorithm}

In the numerically rank deficient situation, the algorithm has to be adapted as detailed in Section~\ref{SS=QRCP_review} with an estimation of the rank $q<p$ and the computation of a new matrix $\widetilde{X}_n \in\R^{q\times n}$ with rank $q$ that replaces $X_n^c$ in Algorithm \ref{ALG:1:QR-full-rank}.
Possible solutions are to scan the values of $\mbox{diag}(|R |)$ from the QR decomposition of $X_n^c$ that are in decreasing order until reaching the first index $q$ such that 
$| R _{q+1,q+1}| < \epsilon | R _{qq}|$, or $| R _{q+1,q+1}| < \epsilon | R _{11}|$ with for example $\epsilon=10^{-8}$. Then a new matrix $\widetilde{X}_n$ is calculated either by truncation or URV decomposition (see Section~\ref{SS=dim_red}).

Let us briefly analyze the computational cost of Algorithm \ref{ALG:1:QR-full-rank}. The QR factorization of the $p\times n$ matrix in Line 1 is executed in $2p^2(n-p/3)$ \emph{flops}.\footnote{A \emph{flop} is one floating point operation - add, multiply, subtract or divide.} The same effort ($2p^2(n-p/3)$ \emph{flops})
is needed to compute the matrix $ Q $; for details see \cite[Section 5.2]{golub-vnl-4}.
The column pivoting (expressed in the permutation matrix $\Pi_1$) precludes full BLAS 3 optimized implementation of the QR factorization, but state of the art libraries such as LAPACK provide efficient implementation that uses block Householder reflectors (\texttt{xGEQP3}). If $n\gg p$ then this can be computed in two steps: first the QR factorization without pivoting and then pivoted QR factorization of the triangular factor. These are implementation details that we omit here.

The costs of Line 2 (data movement),  Line 3 ($O(np)$ \emph{flops} to compute the $q_i$'s) and explicit computation of the matrix $\mbox{Diag}(\sqrt{w((n-1)q_i})_{i=1}^n Q $ in Line 4 are negligible in comparison to the $O(p^2 n)$ cost of Line 1. The matrix $\widetilde{U}_2$ in Line 4 can be computed in $2np^2+11p^3$ flops using the so called R-SVD\footnote{If $n\gg p$ then the computation starts with the QR factorization and then the SVD of the triangular factor is computed. The crossover point depends on the implementation and computing environment.} in the case $n\gg p$; if $n\approx p$ then the cost is $4np^2+8p^3$; see 
\cite[Section 8.6.3]{golub-vnl-4} for more details. The matrix $B$ in Line 5 is computed by solving a triangular system of linear equations at the cost of $p^3$ \emph{flops} (see \cite[Section 3.1]{golub-vnl-4} and \texttt{xTRMM()} in LAPACK). Finally, the essential cost of Line 6 is the matrix multiplication (\texttt{xGEMM} in BLAS 3) that requires $2np^2$ \emph{flops}. 

Altogether, the floating point operation count of Algorithm \ref{ALG:1:QR-full-rank} can be estimated to range from $(53/3)p^3$ (when $n=p$) to $8np^2-(32/3)p^3$ (when $n\gg p$). It should be stressed here that large portion of the computation in the algorithm can be executed using readily available 
LAPACK and BLAS routines that are optimized for the state of the art computer hardware. For comparison, in the same computing environment, the cost of Algorithm \ref{ALG:2:ICS2} can be estimated as $6np^2+26p^3$ (using an estimate $9p^3$ for computing all eigenvalues and eigenvectors of a $p\times p$ real symmetric matrix \cite[Section 8.3]{golub-vnl-4} ).

\begin{remark}
In the numerically rank deficient case (see the rank revealing pivoting in Section~\ref{SS=RRpiv} and Section~\ref{SS=dim_red}), an additional QR factorization of a tall $p\times q$ matrix is computed at the cost of $2(pq^2-q^3/3)$, which in the case $q < p \ll n$ represents a small overhead.
\end{remark}

\section{Applications}\label{S=appli}
In the real world, ill-conditioned data sets are quite common and  arise as soon as the variables are measured in very different units.
For example, industrial data from integrated circuits are made of variables or tests in different units with high accuracy values. Usually, the scales of the measurements are from the pico ($10^{-12}$) or the femto ($10^{-15}$) to the mega ($10^{6}$) or the tera ($10^{12}$). The pico is really common since the electrical capacity is valued in picofarad in the International System.

To illustrate the interest in using the new implementation of ICS, we compared the two algorithms (\textsf{ICSEigen} and \textsf{ICSQR}) on several artificial, and real data sets. We also investigate different applications: clustering, independent component analysis and outlier detection.

To describe how much the data sets suffer from ill-conditioning, we refer to the condition number of the data $X_n$ by $\kappa(X_n)$ which is defined as the ratio of the maximal and the minimal singular value of $X_n$. Usually, the data is said well-conditioned if its condition number is moderate, around $10^k$ with $k=1,2,3$, say, and becomes ill-conditioned when $k$ increases. If $\varepsilon$ is the uncertainty level in the data (including the rounding errors of finite precision computations) then in the well conditioned case $\kappa(X_n)\varepsilon\ll 1$. If the data matrix is singular, the condition number  $\kappa(X_n) = \infty$.

All the following computations are performed with \texttt{R} \cite{R} version 4.1.2. The \textsf{ICSEigen} algorithm corresponds to the \texttt{ics2} function from the \texttt{ICS} \cite{ICS} package and \textsf{ICSQR} is implemented using the \texttt{base} package with the \texttt{qr} and \texttt{eigen} functions. For the \texttt{qr} function, we force the use of LAPACK subroutines to ensure pivoting. The code to reproduce the following applications is available at \url{https://github.com/AuroreAA/NCICS}.

\subsection{Clustering}
 To analyze the behavior of both algorithms, we first focus on ICS as preprocessing step for clustering in the context of an artificial data set as well as real example. The motivation for ICS in this context is that it can recover Fisher's linear discriminant in the case of elliptical mixtures without knowing the class labels and the interesting invariant coordinates belong to components with extreme eigenvalues.

\subsubsection{Mixture of two Gaussian distributions}

Let $Y$ 
be a $p$-variate 
real random vector and assume that the distribution of $Y$ is a mixture of two Gaussian distributions with different location parameters $ \mu_0 = (1,\ldots,1)^\top$ and $ \mu_1 = (\delta, 1,\ldots,1)^\top$, and the same definite positive covariance matrix $\Sigma_W=I_p$:

\begin{equation}
{ Y} \sim (1-\epsilon) \, {\cal N}(\mu_0,\Sigma_W) +  \epsilon \, {\cal N}(\mu_1,\Sigma_W).
 \label{simu}
\end{equation}
Using the mixture model \eqref{simu}, we generate a data set $Y_n$ with $n=10,000$ observations and $p = 4$ variables, with $\epsilon = 0.10$ and $\delta = 6$.

To understand the numerical issues that can happen regarding the conditioning of the data, we rescale the data $Y_n$ and generate data sets $X_n$ with variables in different units.
\begin{equation}
{ X_n} = \diag( c_k) { Y_n}, 
 \label{simu_scale}
\end{equation}

where $c_k$ is a $p$-vector of values by which we multiply $Y_n$ to ensure the condition number of the data set $X_n$ is of $10^{k}$ magnitude, with $ k \in \mathbb{Z}^+$.

\begin{figure}[htbp]
\centering
    \includegraphics[width=0.75\textwidth]{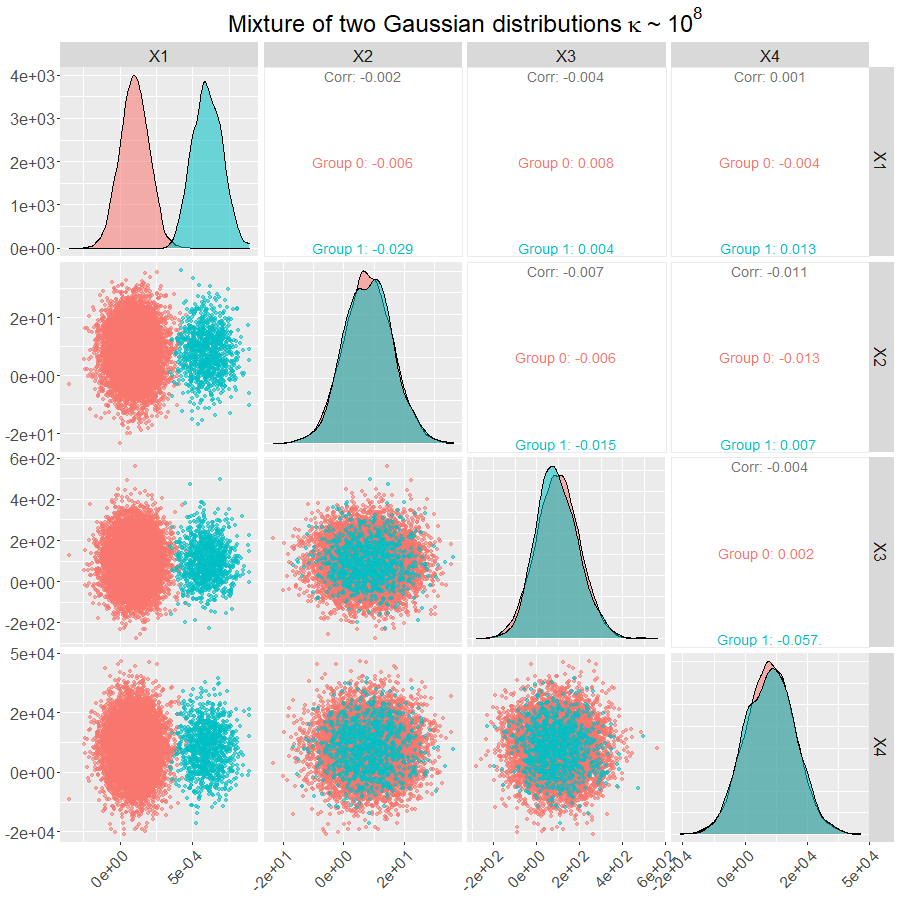}
\caption{Mixture of two Gaussian distributions: 
matrix scatterplots with density estimators on the diagonal and correlations for the scaled data $X_n$ with a condition number around $10^8$.}
\label{fig:simulations_ggpairs}
\end{figure}

Let first take $c_k=(10^{-4}, 10^{1}, 10^{2},  10^{4})$, which corresponds to a condition number around $10^8$. The corresponding scaled data $X_n$ are plotted in Figure~\ref{fig:simulations_ggpairs}. The group separation is clearly highlighted on the first variable. 
Note that the plots are identical for the unscaled data $Y_n$ and for any other scaled data, up to the units on the axes. 

\begin{figure}[htbp]
\centering
    \includegraphics[width=1\textwidth]{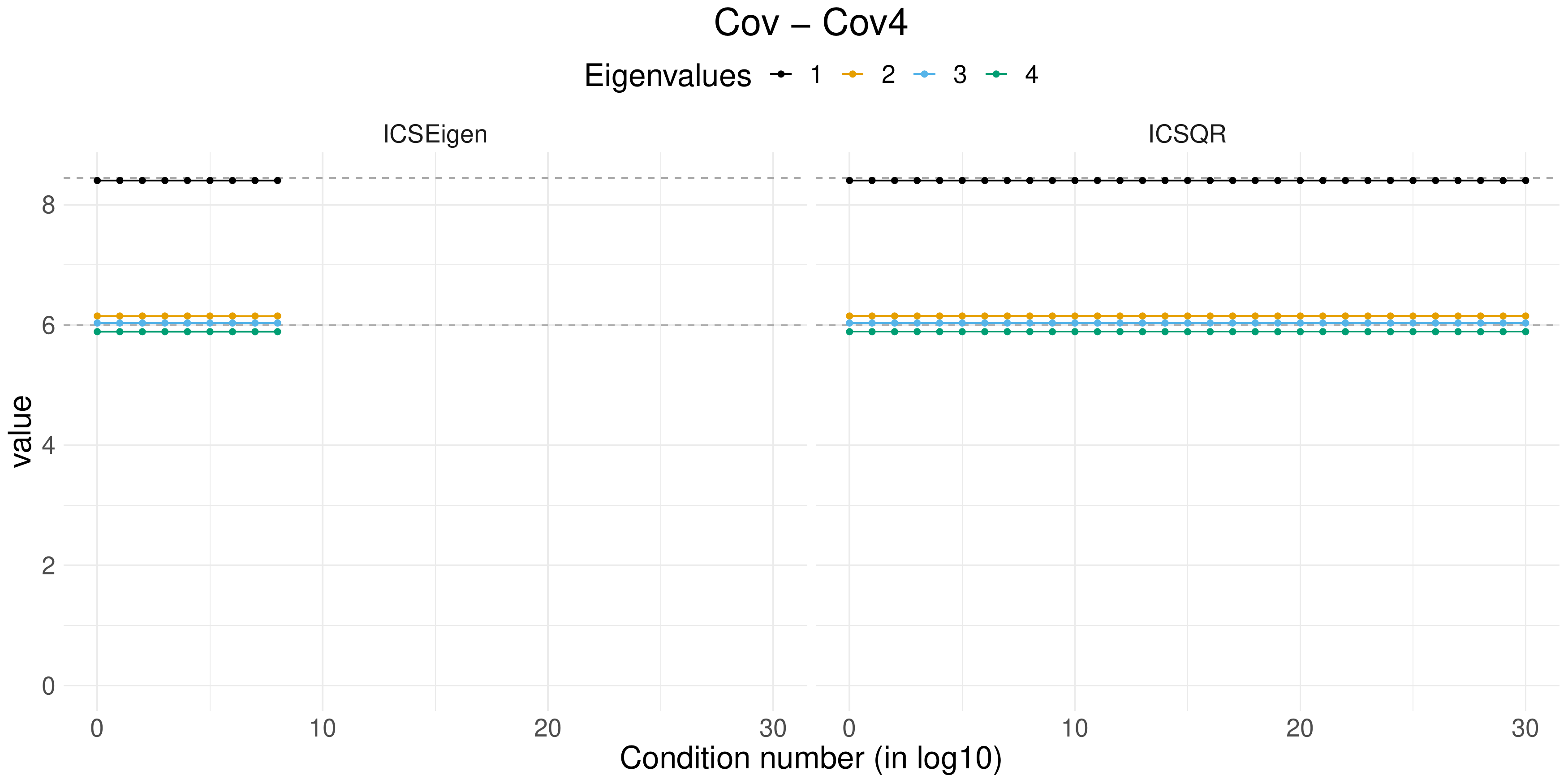}
      \includegraphics[width=1\textwidth]{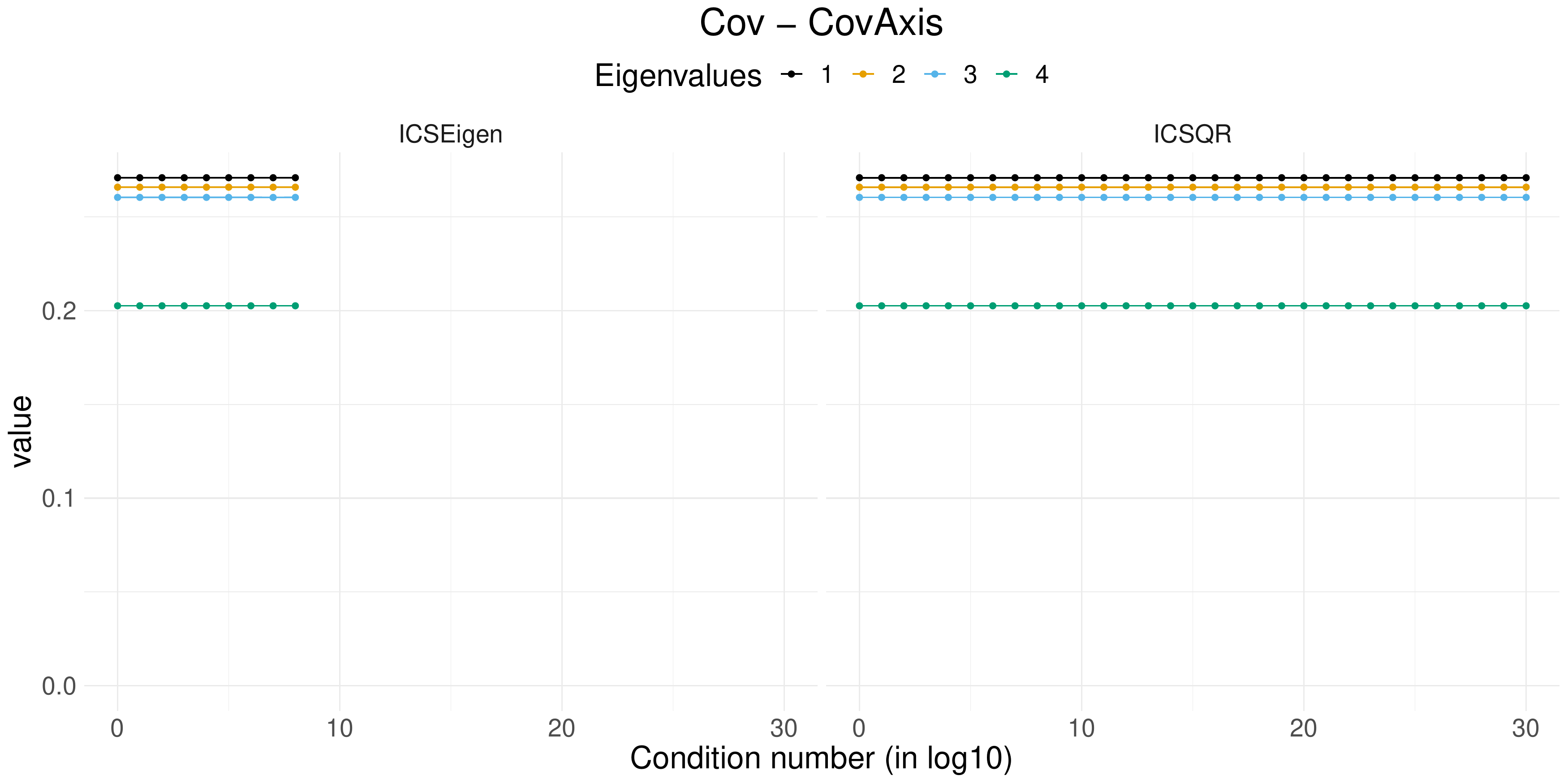}
\caption{Mixture of two Gaussian distributions: comparison of the eigenvalues derived by the two algorithms when the condition number of the data $X_n$ varies from $k=0,\ldots,30$. Each row corresponds to a different combination of scatters:  $\cov$ - $\cov_{4}$ (first row) and $\cov$-$\cov$Axis (second row). Each column refers to an algorithm: \textsf{ICSEigen} (first column), \textsf{ICSQR} (second column).}
\label{fig:simulations_eigenvalues}
\end{figure}

Let us now generate different scale vectors $c_k$ such that the condition number of $X_n$ vary between $10^0$ and $10^{30}$.
We compare the eigenvalues computed by the two algorithms: \textsf{ICSEigen} and \textsf{ICSQR}, for different combinations of scatter matrices: $\cov$ and $\cov_{w}$ with $w(d) = d^{\alpha}$ and $\alpha = -1,-0.5,0.5,1$. In Figure~\ref{fig:simulations_eigenvalues}, we focus only on the cases with $\alpha =1$ and $\alpha = -1$ which correspond to the well-known scatters $\cov_4$ (used in FOBI), and $\cov$Axis (used in PAA). The results with other values for $\alpha$ are similar and are not presented here.

The main remark is that \textsf{ICSEigen} is not able to perform the computation of ICS as soon as the condition number of the data is higher than $10^8$, and terminates with an error mentioning that the system is computationally singular. 
\textsf{ICSQR} does not present such issues and the four eigenvalues continue to be stable over the increase of $k$. In addition, for the combination of scatters $\cov$-$\cov_4,$ we can also compare obtained eigenvalues 
to the known theoretical ones, 
represented by dashed lines in Figure~\ref{fig:simulations_eigenvalues}, as the exact computation is trivial in this case.  As expected, only one eigenvalue is clearly different from the others, since we have simulated a data set containing only two groups. The three other eigenvalues are not exactly equal but are stable and close to the theoretical ones. The conclusions are similar for the combination of scatters $\cov$-$\cov$Axis, except that this time the eigenvalue of interest is the last one. 

\subsubsection{Crabs data}
\begin{figure}[htbp]
\centering
    \includegraphics[width=0.49\textwidth]{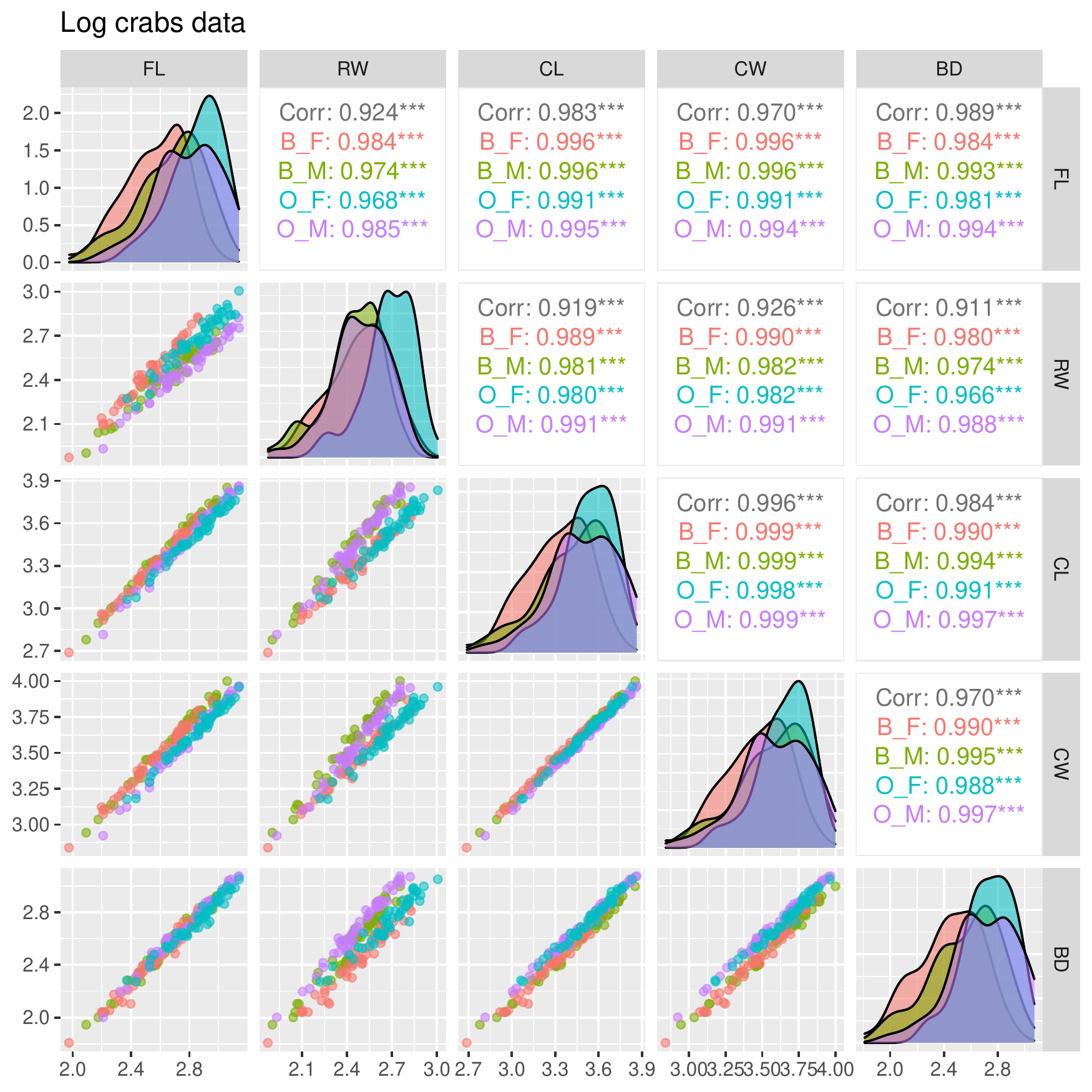}
    \includegraphics[width=0.49\textwidth]{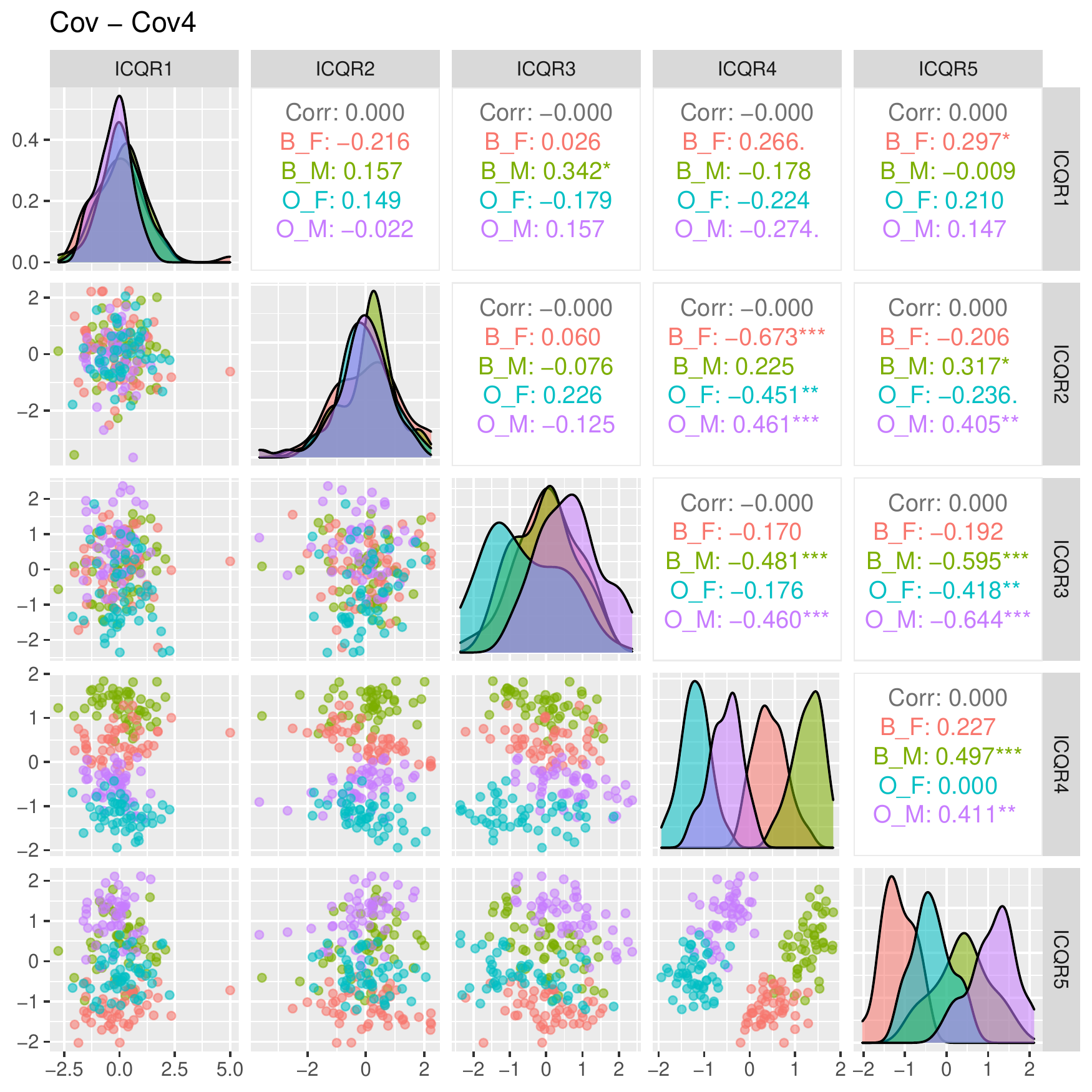}
\caption{Crabs data: the left panel is the scatterplot matrix of the initial data (after a logarithmic transformation). The right panel is the scatterplot matrix of the invariant components obtained with \textsf{ICSQR} for the the combination of scatters $\cov$ and $\cov_4$ and the scaled data with a condition number around $10^9$.}
\label{fig:crabs_log_ggpairs}
\end{figure}

Another example is the well-known crabs data set \cite{campbell1974multivariate}, containing 200 observations on five continuous variables regarding morphological measurements of crabs. As it is quite common  we prepare the data by taking log-transformations of all variables. There are two species of both sexes and all groups are balanced (50 in each), yielding therefore 4 equally sized groups. The data set is represented in the left panel of Figure~\ref{fig:crabs_log_ggpairs}.

Here this data set is well-conditioned ($k \approx 3$) and so using \textsf{ICSEigen} or \textsf{ICSQR} gives exactly the same eigenvalues and the same components up to inaccuracies of order of magnitude around $10^{-12}$. Both algorithms highlight the four groups in the first (or last) two components depending on the combination of scatters used. Each time, one component makes a clear separation by species and the other one by sexes.
However, if we scale the data set to have different units for each variable, \textsf{ICSEigen} no longer can be performed as soon as the condition number of the data is higher than $10^8$, like in the artificial data set. The right panel of Figure~\ref{fig:crabs_log_ggpairs} presents the scatterplot matrix of the invariant components obtained with \textsf{ICSQR} and the combination of scatters $\cov$ and $\cov_4$ for a scaled version of the crabs data set with a condition number around $10^9$. As expected, \textsf{ICSQR} can be performed even in this context and the results are stable and comparable 
to the ones on the initial data: the four groups are clearly identified by the last two components.

\subsection{Independent Component Analysis}

Independent component analysis (ICA) is a model-based multivariate method where it is assumed that the observed data is a linear mixture of non-Gaussian independent components (at most one Gaussian component is allowed). The aim of ICA is then to recover the latent independent components, and there are many methods for their extraction, most of which are being based either on projection pursuit or on the simultaneous use of two scatter functionals (the ICS approach).  It can be shown that in the ICA model the diagonal elements of $\bo D$ in~\eqref{eq:eqICS} correspond to kurtosis measures of the latent components (in this special case with respect to $\cov-\cov_w$ combination of scatters); see \cite{miettinen2014fourth} for more insight. Therefore, if all independent components have distinct kurtosis values w.r.t the chosen combination of scatters, it can be shown that ICS solves the ICA problem, provided the chosen combination of scatters fulfills some additional assumptions (for details see for example \cite{NordhausenTyler2015}, \cite{Radojicic2020ngcaTest}). In that case, the invariant coordinates correspond to the original latent independent components, up to signs and order. For a general review on ICA see for example \cite{NordhausenOja2018}.

For the demonstration purposes we simulate data from the ICA model where the latent components follow a normal, a $t_5$, a uniform and a Laplace distribution, always centered and standardized to unit variance. In this setup, all scatter combinations discussed for ICS in this paper find the latent independent components. Due to the affine equivariance of ICS, the choice of the mixing matrix  (inverse of the unmixing matrix) is immaterial and we use for simplicity a diagonal matrix that in turn can be used to rescale the data to obtain data set with an arbitrary condition number. 

\begin{figure}[htbp]
\centering
    \includegraphics[width=1\textwidth]{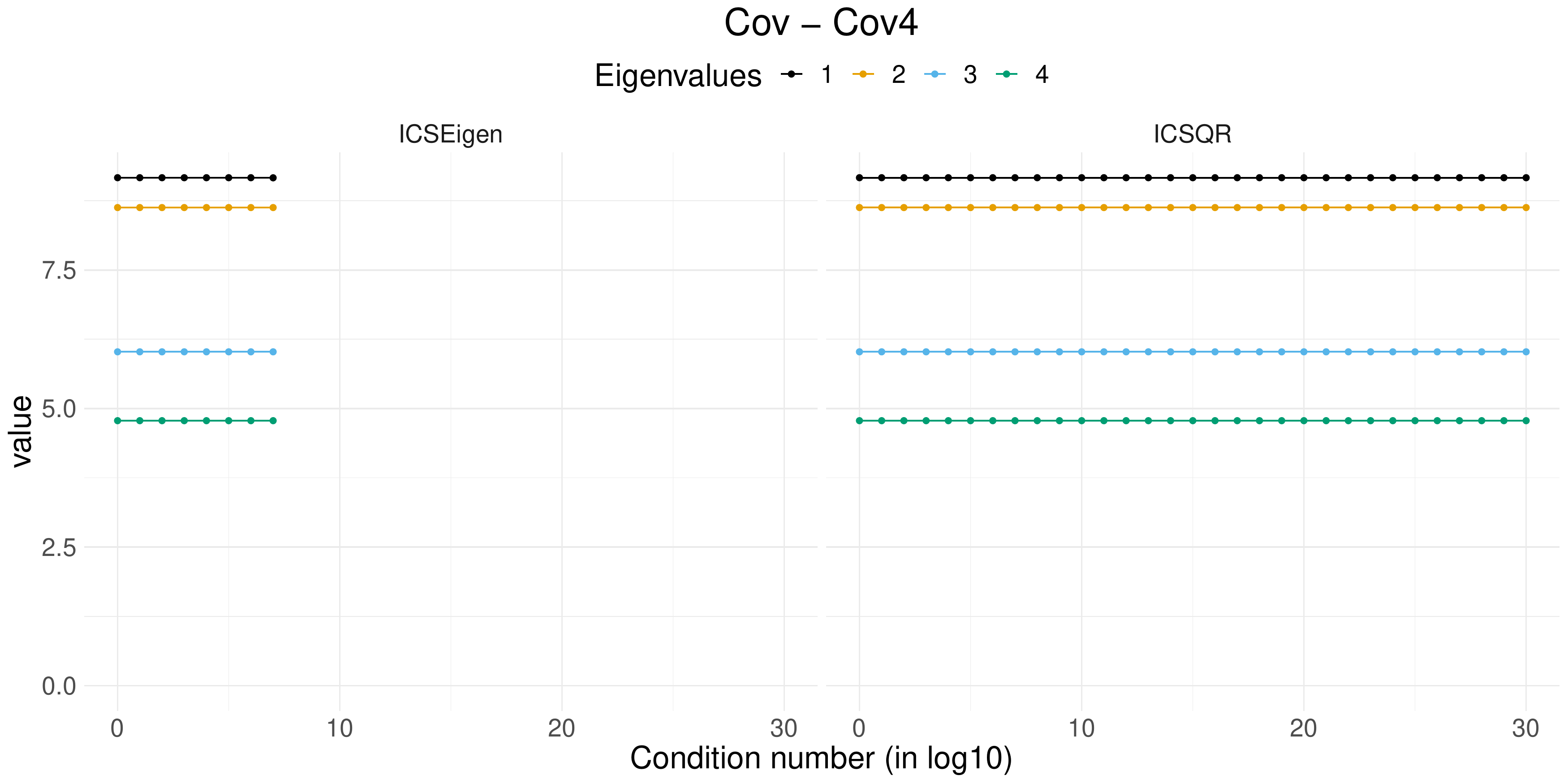}
     \includegraphics[width=1\textwidth]{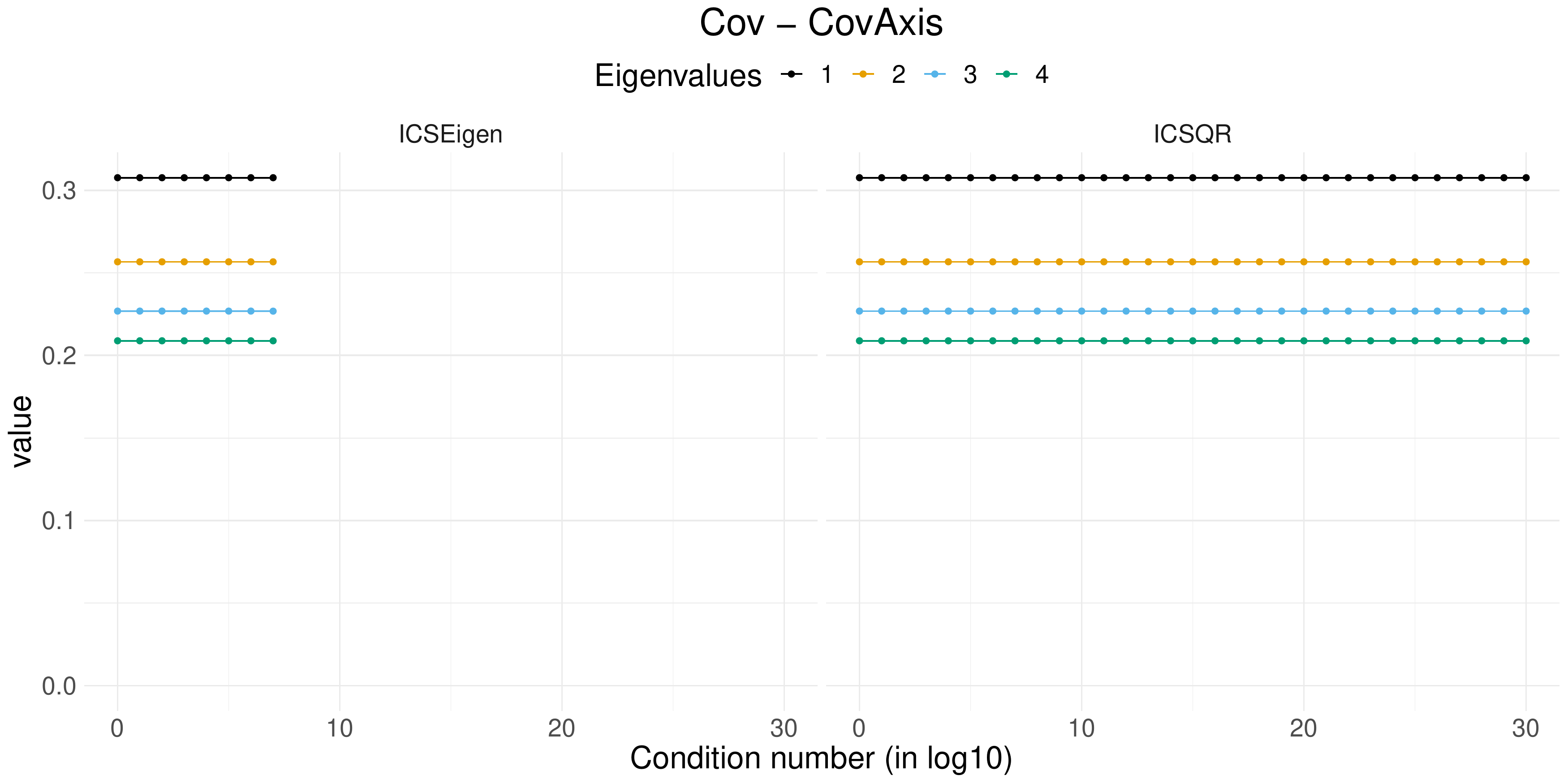}
\caption{ICA example: Comparison of the eigenvalues derived by the two algorithms when the condition number of
the data  varies from $k = 0,\ldots, 30$. Each row corresponds to a different combination of scatters: COV- COV4 (first row) and COV-COVAxis (second row). Each column refers to an algorithm: \textsf{ICSEigen} (first
column), \textsf{ICSQR} (second column). }
\label{fig:simulations_ICA_eigenvalues}
\end{figure}

Figure~\ref{fig:simulations_ICA_eigenvalues} plots the ICS eigenvalues of \textsf{ICSEigen} and \textsf{ICSQR} against the condition number in the case of $\cov$-$\cov_4$ and $\cov$-$\cov$Axis.

Figure~\ref{fig:simulations_ICA_eigenvalues} clearly shows that, as in the previous cases, \textsf{ICSEigen} fails 
when the condition number gets too large, while \textsf{ICSQR} keeps producing 
stable results (eigenvalues) when data is even more ill-conditioned.

\subsection{Outlier Detection}

Finally, a well-known application of ICS is  outlier detection and \cite{archimbaud2018CSDA, archimbaud2018Rjournal} propose a general procedure using ICS for that purpose in the case of a small proportion of outliers.  As explained in \cite{archimbaud2018CSDA}, only the first components are then of interest for the combination of scatters $\cov$ and $\cov_4$. The choice of the number of components to retain is not the issue investigated here, so we focus only on the first component for two real data sets coming form an industrial quality control background. The  anonymized data sets are available at \url{https://github.com/AuroreAA/NCICS}.

\subsubsection{HTP3: nearly singular industrial data}

\begin{figure}[htbp]
\centering
    \includegraphics[width=1\textwidth]{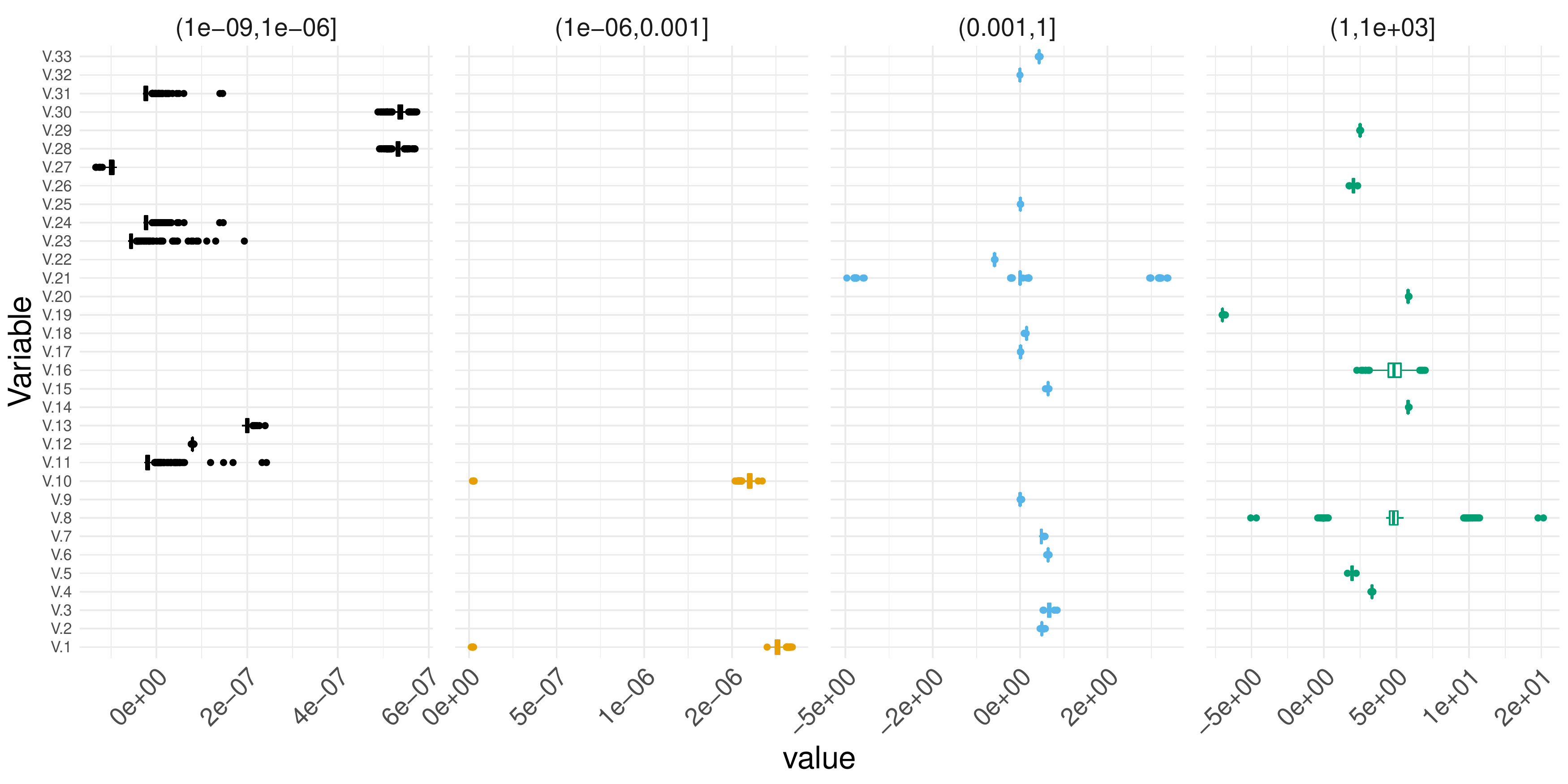}
\caption{HTP3 data set: boxplots of the 33 variables, presented in 4 different groups regarding the scale of their absolute mean.}
\label{fig:HTP3_box}
\end{figure}

We first consider the data called HTP3 which contains 371 high-tech parts designed for consumer products and characterized by 33 tests as illustrated in Figure~\ref{fig:HTP3_box}. 
These tests are performed to ensure a high quality of the production and are anonymized here. All  371 parts were considered functional and reached the market. However, part 32 showed defects in use and was returned to the manufacturer by its buyer.  The question is thus if this part could have been identified as malfunctioning based on the performed tests.

\begin{figure}[htbp]
\centering
    \includegraphics[width=0.75\textwidth]{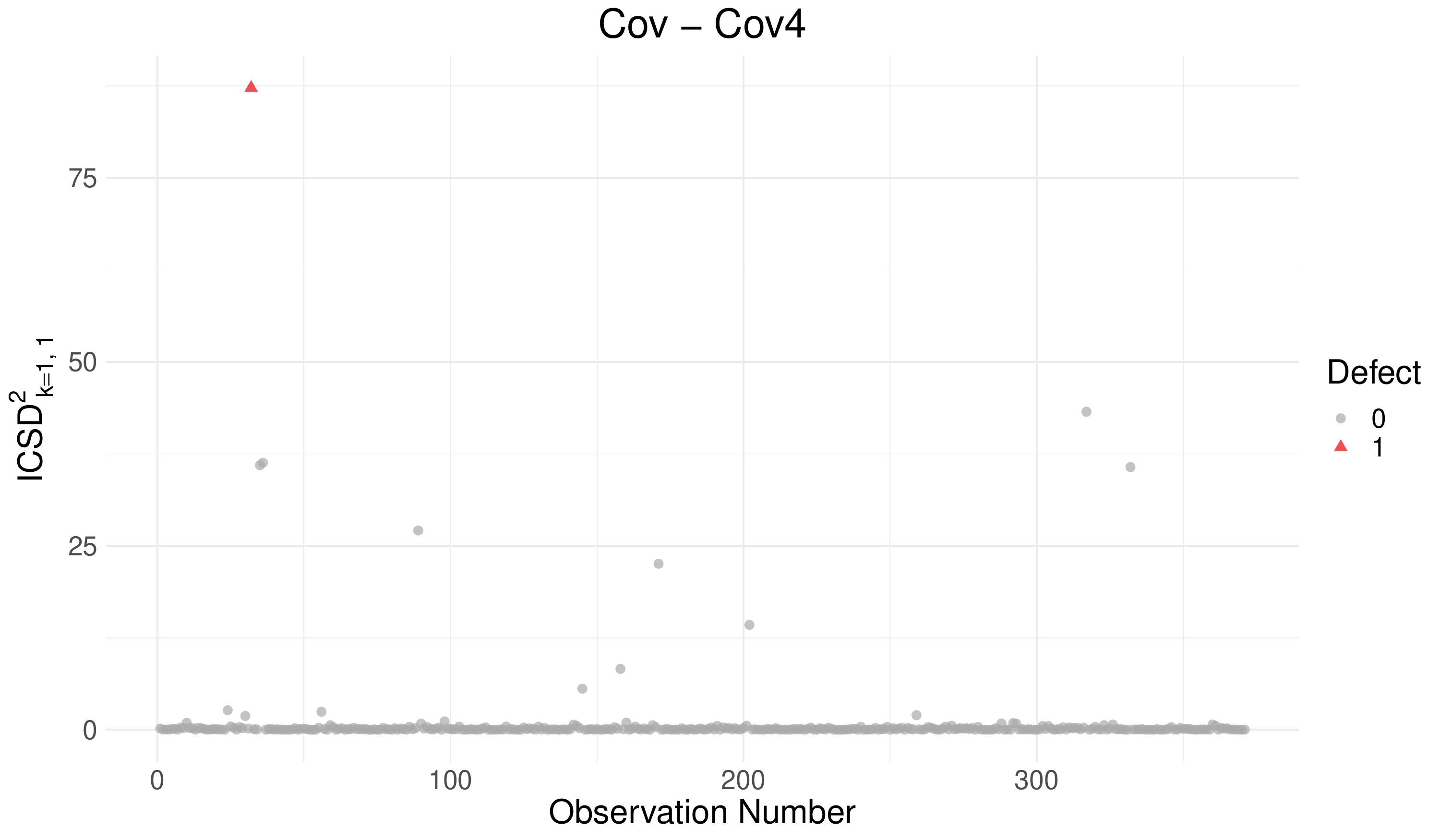}
\caption{HTP3 dataset: ICS distances computed with one component for the combination of scatters $\cov$ and $\cov_4$. The defective part is represented in red.}
\label{fig:HTP3_ICSD}
\end{figure}

The tests are visualized in Figure~\ref{fig:HTP3_box}, where it is clearly visible that the measurements have completely different units: nine variables have their absolute mean between $(10^{-9},10^{-6}]$, two between $(10^{-6},10^{-3}]$, 13 between $(10^{-3},10^{0}]$ and 9 between  $(10^{0},10^{3}]$, yielding condition number for the data set of around $10^{10}$. For example computing the eigenvalues of the covariance matrix for the data set in R yields two negative eigenvalues indicating that we are close to singularity;
 [1]  1.029573e+01  2.230258e+00  5.892883e-01  1.685965e-02  3.064711e-03
 ...
 [31]  8.811646e-19 -1.010746e-17 -8.841896e-16.
Consequently also the \textsf{ICSEigen} algorithm terminates with an error. On the contrary the \textsf{ICSQR} algorithm does not encounter any issues and can compute all the eigenvalues, the first one begins around 100 and the last one around 30, for the $\cov-\cov_4$ combination of scatters.  As explained in \cite{archimbaud2018CSDA} using the so-called squared ICS distances, denoted ICSD$^2$, of selected components should then reveal outliers. Figure~\ref{fig:HTP3_ICSD} shows these distance based on the first invariant component. And indeed, the defective part 32, represented by a red triangle, has the highest ICS distance indicating that it is different from the majority of the observations.


\subsubsection{HTP2: collinear industrial data}
\begin{figure}[htbp]
\centering
    \includegraphics[width=1\textwidth]{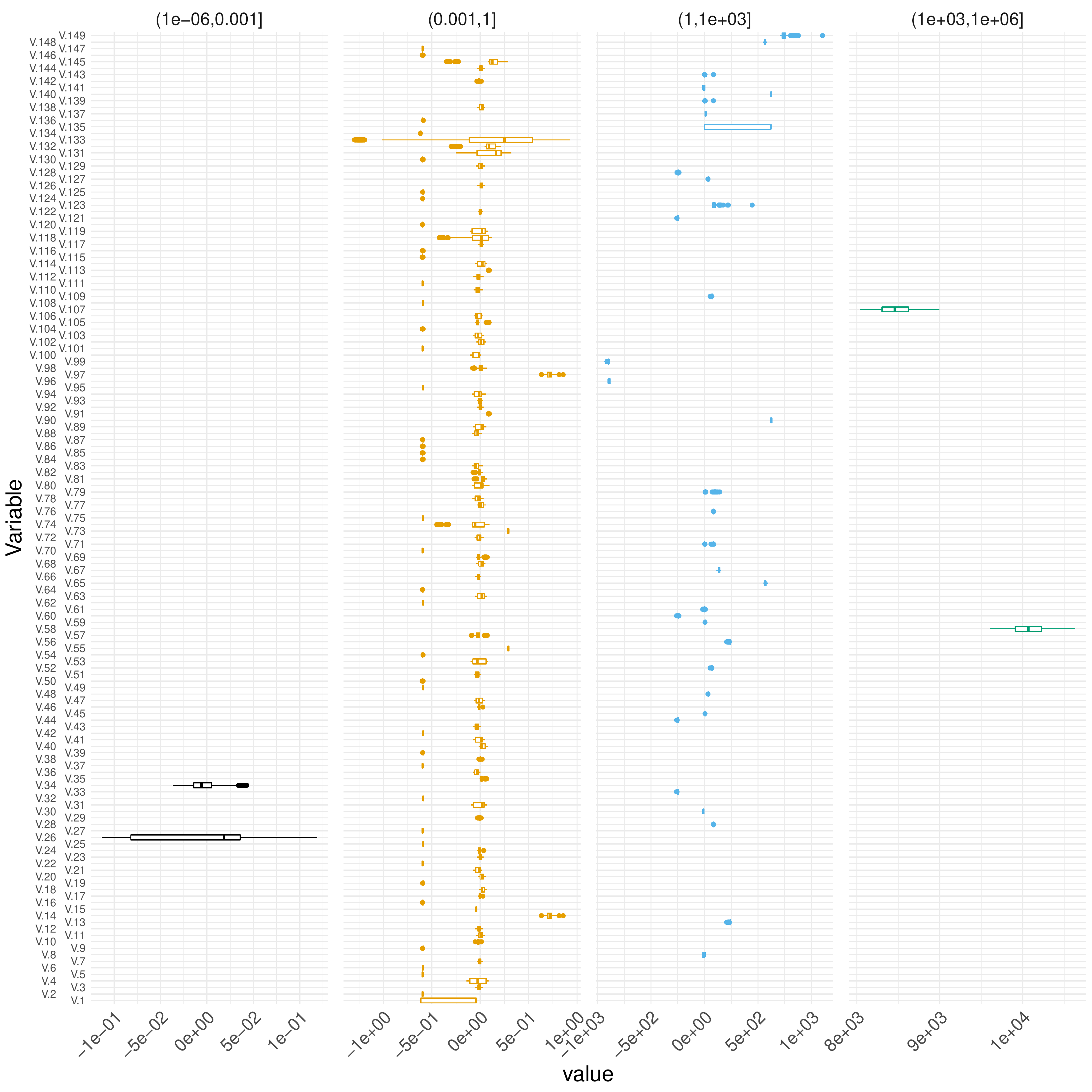}
\caption{HTP2 data set: boxplots of the 149 variables, presented in 4 different groups regarding the scale of their absolute mean.}
\label{fig:HTP2_box}
\end{figure}

While the data set HTP3 had quite many observations relative to the number of performed tests, data set HTP2, which has a similar context as HTP3, contains 149 tests for 457 high-tech parts where the known defect is 
the part 28. The condition number for this data set is around $10^{22}$  indicating again ill-conditioning that we suspect is due to collinearity between the variables. Again the measurement units for the tests are in rather different units as illustrated in Figure~\ref{fig:HTP2_box} and the eigenvalues of the covariance matrix are in three instances negative, again prohibiting the use of \textsf{ICSEigen}.
Contrary to the previous cases, the error message is ``the system is exactly singular" instead of ``computationally singular". The data are singular and even the \textsf{ICSQR} algorithm fails because the problem is not a numerical one.

A common practical approach in such a case is often to reduce first the dimension of the data and then apply the method of interest to the reduced data. This for example can be done by projecting the data using the right-singular vectors of an SVD associated with non-zero singular values.
The critical question is then to decide which singular values are actually zero? There seems not to be any definite rule but a consensus seems to be to use the relative criterion: $ \frac{\rho_i}{\rho_{\max}} \leq \text{tolerance} $, where however  $\text{tolerance}$ needs to be specified. 
We use a tolerance level commonly used in R, i.e., $\text{tolerance}= \max(n,p) \epsilon_m$ with $\epsilon_m$ being the accuracy of the machine (in R, $\epsilon_m \approx  2.22e^{-16}$).


Based on this procedure, the base R  function \texttt{svd} gives an estimated rank for HTP2 of 141 with the first eigenvalue equal to $2.8e^{+05}$ and the last to $4.8e^{-04}$. But even after this reduction step, \textsf{ICSEigen} fails because the system is now `computationally singular' while \textsf{ICSQR} works. To ensure the most accurate results, we thus suggest the approach from Section~\ref{SS=QRCP_review} that makes use of a pivoted QR factorization.

\begin{figure}[htbp]
\centering
    \includegraphics[width=0.75\textwidth]{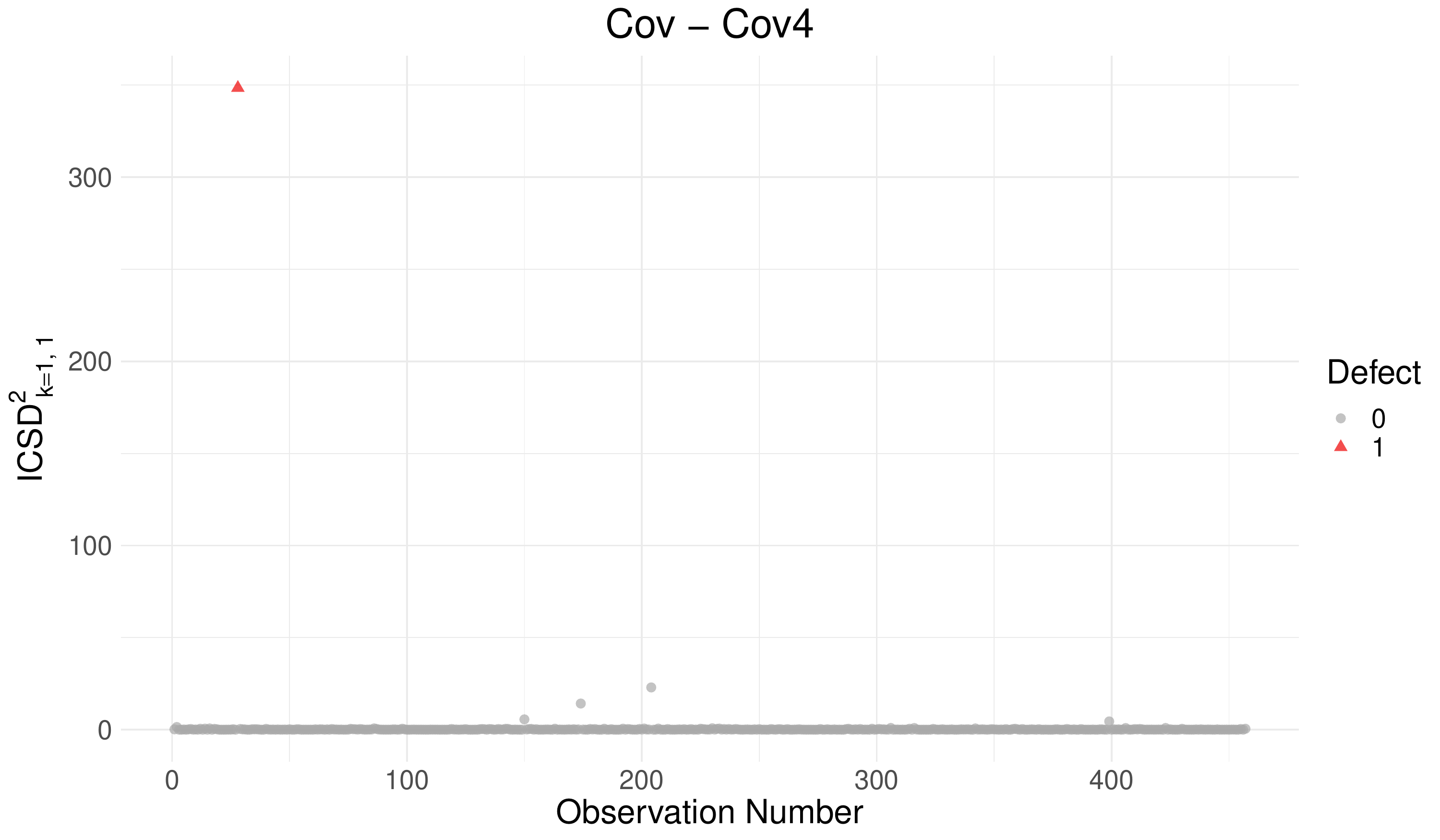}
\caption{HTP2 data set: ICS distances computed with one component for the combination of scatters $\cov$ and $\cov_4$ on the reduced data of rank 141 with the \textsf{ICSQR} algorithm. The defective part is represented in red.}
\label{fig:HTP2_ICSD}
\end{figure}

This preprocessing step also estimates the rank of the data to be 141, using the same threshold as previously, with the first value being  $2.7e^{+02}$ and the last one being $4.9e^{-07}$. A consequence of such preprocessing is that the data is being permuted in a way that ensures higher numerical stability of the subsequent methods. More precisely, data is permuted and reduced ensuring that we keep a well-conditioned part of the data set, making it possible to perform ICS using both algorithms, \textsf{ICSEigen} and \textsf{ICSQR}. Using then again the scatter combination $\cov$ and $\cov_4$ and computing the ICS distances based on the first component reveals clearly 
28 as 
an outlier as shown in Figure~\ref{fig:HTP2_ICSD}.

\section{Conclusion}\label{S=conclu}

ICS is an increasingly popular multivariate method with many application areas. So far mainly theoretical properties were considered in the literature, while the computational aspects were neglected. In one of the main areas of applications of ICS, outlier detection in industrial quality control, it was however observed that there are often computational problems in practical use of ICS. Therefore, for a broad general group of scatter combinations in ICS, we examined numerical properties of the estimation method and suggest a new algorithm resolving many of the numerical issues, providing therefore a clear improvement.
The method makes use of the QR factorization and allows to avoid in a clever way the calculation of the inverse and the square root of the inverse of the covariance matrix. The importance of the column and row pivoting is also discussed in detail. While the row pivoting may help to reduce the the errors of the finite precision arithmetic, the column pivoting leads also to a rank revealing procedure that is useful when the data are exactly singular as it is the case in high dimension ($p>n$). Several examples illustrate the advantage of using the new algorithm especially in situations where the measurement units highly differ between variables.

The suggested algorithm is constructed for the $\cov-\cov_w$ class of scatter combinations, that includes for example some widely used combinations like FOBI and PAA, as special cases. However, the newly established algorithm cannot be applied to arbitrary scatter combinations.  In future research we will consider if such algorithmic improvements are possible for other scatter combinations.



\section*{Acknowledgments}
This work was partly supported by a grant of the Dutch Research Council (NWO, research program Vidi, project number VI.Vidi.195.141), by the French \textit{Agence Nationale de la Recherche} through the Investments for the Future (Investissements d'Avenir) program, grant ANR-17-EURE-0010,  by the Croatian Science Foundation (CSF) grant IP-2019-04-6268, and by the Austrian Science Fund P31881-N32.

\bibliographystyle{unsrt}  
\bibliography{numICS}

\end{document}